\documentclass[a4paper,fleqn,usenatbib,useAMS]{mnras}

\usepackage{graphicx}	% Including figure files
\usepackage{amsmath}	% Advanced maths commands
\usepackage{amssymb}	% Extra maths symbols
\usepackage{natbib}
\usepackage[flushleft]{threeparttable}
\usepackage{pdflscape}	% Landscape pages
\usepackage[T1]{fontenc}
\usepackage{ae,aecompl}

\title[Extended molecular gas at high redshift]{Molecular gas on large circumgalactic scales at z=3.47}

\author[M. Ginolfi et al.]{
M. Ginolfi,$^{1,2,3}$\thanks{E-mail: mg791@mrao.cam.ac.uk}
R. Maiolino,$^{2,3}$
T. Nagao,$^{4}$
S. Carniani,$^{2,3}$
F. Belfiore,$^{2,3}$
G. Cresci,$^{5}$ \newauthor
B. Hatsukade,$^{6}$
F. Mannucci,$^{5}$
A. Marconi,$^{5,7}$
A. Pallottini,$^{2,3,8}$
R. Schneider,$^{1}$
P. Santini$^{1}$
\\
$^{1}$INAF/Osservatorio Astronomico di Roma, Via di Frascati 33, 00040 Monte Porzio Catone, Italy\\
$^{2}$Cavendish Laboratory, University of Cambridge, 19 J.J Thomson Ave., Cambridge CB3 0HE, UK\\
$^{3}$Kavli Institute for Cosmology, University of Cambridge, Madingley Road, Cambridge CB3 0HA, UK\\
$^{4}$Research Centre for Space and Cosmic Evolution, Ehime University, 790-8577 Ehime, Japan\\
$^{5}$ INAF/Osservatorio Astrofisico di Arcetri, Largo E. Fermi 5, Firenze, Italy\\
$^{6}$ National Astronomical Observatory of Japan, 2-21-1 Osawa, Mitaka, 181-8588 Tokyo, Japan\\
$^{7}$ Dipartimento di Fisica e Astronomia, Universita' di Firenze, Via G. Sansone 1, Sesto Fiorentino, Italy\\
$^{8}$ Scuola Normale Superiore, Piazza dei Cavalieri 7, I-56126 Pisa, Italy
}

\date{Accepted XXX. Received YYY; in original form ZZZ}

\pubyear{2016}

\begin{document}
\label{firstpage}
\pagerange{\pageref{firstpage}--\pageref{lastpage}}
\maketitle

\begin{abstract}
We report ALMA observations of the most massive (star forming) galaxy in the redshift range $\rm 3<z<4$ within the whole GOODS-S field.
We detect a large elongated structure of molecular gas around the massive primeval galaxy, traced by the CO(4-3) emission, and extended over 40 kpc. 
We infer a mass of the large gaseous structure of $\rm M_{gas}\sim2-6 \times 10^{11} M_{\odot}$. 
About 60\% of this mass is not directly associated with either the central galaxy or its two lower mass satellites. 
The CO extended structure is also detected in continuum thermal emission.
The kinematics of the molecular gas shows the presence of different components, which cannot be ascribed to simple rotation.
Furthermore, on even larger scales, we detect nine additional CO systems within a radius of 250~kpc from the massive galaxy and mostly distributed in the same direction as the CO elongated structure found in the central 40 kpc. The stacked  images of these CO systems show detections in the thermal continuum and in the X-rays, suggesting that these systems are forming stars at a rate of 30--120~$\rm M_{\odot} yr^{-1}$.
We suggest that the extended gas structure, combined with its kinematic properties, and the gas rich star forming systems detected on larger scales, are tracing the inner and densest regions of large scale accreting streams, feeding the central massive galaxy.
These results corroborate models of galaxy formation, in which accreting streams are clumpy and undergo some star formation (hence enriching the streams with metals) even before accreting onto the central galaxy.
\end{abstract}


%%%%%%%%%%%%%%%%%%%%%%%%%%%%%%%%%%%%%%%%%%%%%%%%%%

%%%%%%%%%%%%%%%%% BODY OF PAPER %%%%%%%%%%%%%%%%%%

\section{Introduction}

Modern theories of galaxy evolution and cosmological numerical simulations predict that massive galaxies, especially in the early stages of their formation, accrete large amount of gas from the intergalactic medium (IGM), mostly through large scale filamentary structures funnelling gas into the galaxy dark matter (DM) halo towards the disc (\citealp{Dekel2009}; \citealp{Silk2012}; \citealp{Genel2012}).  \newline
Such gas replenishment has to occur continuously throughout the cosmic time and more efficiently at early epochs, to sustain the intense star formation observed in primeval systems (\citealp{Dekel2009}; \citealp{SanchezAlmeida2014}; \citealp{Keres2009}). 
These theoretical claims are observationally supported by the finding that high redshift star forming galaxies are characterized by gas depletion timescales of the order of, or shorter than, 1~Gyr, hence requiring that fresh gas must be supplied at high rate across a significant fraction of the  galaxy lifetime, at least while populating the so-called main sequence \citep[e.g.][]{Tacconi2013,Genzel2015,Scoville2016}. 
The need for prolonged star formation timescales sustained by gas accretion is also inferred from studies of the chemical enrichment of spiral galaxies (e.g. \citealp{Tinsley1981}; \citealp{Matteucci1989}; \citealp{Matteucci2014}).  

The additional gas needed to support star formation can be provided by the IGM, feeding massive primeval galaxies via gas accretion through cosmic streams (\citealp{Sancisi2008}; \citealp{Fraternali2014}). 
Theoretical models and numerical simulation have thoroughly investigated the role of gas accretion in galaxy evolution.  
The standard view of \textquoteleft hot accretion mode\textquoteright\, has persisted for years, arguing that galaxies form out of collapsing, virialized gas that forms an hot halo and then slowly cools, and settles in a disk (\citealp{Rees1977}; \citealp{White1978}; \citealp{Fall1980}). 
This view has been substantially modified in the past decade and the new paradigm of
\textquoteleft cold accretion mode\textquoteright\, predicts that, at early cosmic epochs,
the majority of  gas enters the DM halo in the cold phase along filaments feeding the disk
growth, without significant shock heating, at least at $\rm z>3.5$ (\citealp{Birnboim2003}; \citealp{Dekel2006}; \citealp{Keres2009}; \citealp{Ocvirk2008}; \citealp{Brooks2009}; \citealp{Stewart2011}).
It should be noted that such models expect that cold streams are halted in massive galaxies ($\rm M_{\star}>10^{11}~M_{\odot}$) at $\rm z<3-3.5$ as a consequence of halo shock heating ({\citealp{Dekel2009}; \citealp{Cattaneo2006}}).  

Our current understanding of cold gas inflows is largely based only on such models, as observational studies have been severely limited by the difficulty of tracing such cold gas accretion. \newline
Attempts have been made in tracing the distribution of circumgalactic gas around high redshift galaxies by looking at distinct kinematic signatures in absorption systems (along the line of sight of background quasars) produced by accreting material (e.g. \citealp{Bouche2013}; \citealp{Prochaska2014}). 
Unfortunately, this approach is limited by the sparseness of aligned quasar-high redshift galaxies pairs and by the fact that only one line of sight is usable, which results in degeneracies on the inferred CGM distribution.
Another powerful tool to observe the extended gas reservoir surrounding high redshift galaxies is through direct imaging of Lyman-$\alpha$ (Ly$\alpha$) emission due to recombination radiation following photoionization (sometimes refereed to as  \textquoteleft fluorescence\textquoteright\,) powered by powerful ultraviolet (UV) sources (\citealp{Lowenthal1990};  \citealp{Haardt1996};\citealp{Bunker1998}; \citealp{Cantalupo2005}; \citealp{Geach2009}; \citealp{Kollmeier2010}).  
Successful ultra-deep observations of hyper-luminous quasars at $\rm z>2$ have indeed revealed giant Ly$\rm \alpha$ haloes with projected linear sizes larger than 100 kpc (\citealp{Cantalupo2012}; \citealp{Hennawi2015}; \citealp{Cantalupo2014}; \citealp{Borisova2016}). 
Recently, extensive integral field spectroscopic surveys have lead to the first detection of the 2-dimensional distribution of spatially extended Ly$\rm \alpha$ emission also around low mass ($\rm\sim10^8-10^9\, M_{\odot}$) star-forming galaxies at $\rm z>3$ (\citealp{Wisotzki2016}; \citealp{Patricio2016}; \citealp{Vanzella2016}), revealing that Ly$\rm \alpha$ haloes appear to be ubiquitous even among \textquoteleft normal\textquoteright\, galaxies at high redshift.  \newline
Although such results seem to confirm what predicted by models, revealing the extended gaseous structures expected to exist around primeval galaxies, the physical properties and the nature of the cosmic accreting gas expected by theories are still debated. 
Indeed, observations of Ly$\rm \alpha$ haloes are only able to offer limited insight into the nature of the CGM because of (1) the possibility of tracing only the warm ionized phase of the gas surrounding primeval galaxies (which is generally a small fraction of total), (2) the degeneracies affecting the physical parameters estimated from radiative transfer modelling of Ly$\rm \alpha$ emission and (3) the inability to infer detailed information about the kinematics of the 3-dimensional spatially extended gas, both because of Ly$\alpha$ absorption and the resonant nature of the line.
%%%%%%%%%%%%%%%%%%%%%%%%%%%%%%%%%%%%%%%%%%%%%
\begin{table*}
	\centering
	\caption{Summary of the properties of Candels-5001}
	\label{tab:candels_only}
	\begin{threeparttable}
		\begin{tabular}{ccccccccc} % four columns, alignment for each
			\hline
			RA &DEC &redshift& 12+log(O/H)\tnote{a}& log($\rm M_{star}/M_{\odot}$)\tnote{b}&F(CO$_{4-3}$)\tnote{c}&log($\rm M_{H_2}/M_{\odot}$)\tnote{c}& $\rm F_{cont.}(103GHz)$ &log(SFR) \tnote{d}\\
			J2000&J2000&&&&Jy $\rm km\,s^{-1}$&& $\rm \mu Jy$&$\rm M_{\odot}\, yr^{-1}$\\ 
			\hline 
			03:32:23:336&-27.51.56.862 &3.473 &$8.41^{+0.10}_{-0.10}$&$10.27^{+0.38}_{-0.11}$&$0.13\pm 0.02$&$11.09^{+0.17}_{-0.25}$&$36.1\pm 9.6$&$2.33 \pm 0.31$\\ 
			\hline
		\end{tabular}
		\begin{tablenotes}
			\item[a] Obtained using the empirical calibration derived by \cite{Curti2016}.
			\item[b] \cite{Santini2015}.
			\item[c] This work.
			\item[d] From the extinction-corrected H$\beta$ emission \citep{Troncoso2014}.
		\end{tablenotes}
	\end{threeparttable}
\end{table*}
%%%%%%%%%%%%%%%%%%%%%%%%%%%%%%%%%%%%%%%%%%%%%

Such limits allow observations to only partly confirm models, leaving the observational evidence of  gas cosmic streams feeding high redshift galaxies still sparse and debated. 
However, due to the increasing accuracy of numerical cosmological simulations, the theoretical scenario has substantially evolved over the last few years. 
While initial theories were predicting inflows of nearly pristine gas (\citealp{Birnboim2003}; \citealp{Keres2005}; \citealp{Dekel2006}), recent simulations have shown that cooling and gravitational collapse may happen in gas streams, possibly leading to star formation within them, resulting in metal enrichment prior to delivery onto the massive galaxy (\citealp{Nelson2016}; \citealp{Ceverino2016}; \citealp{Pallottini2014, Pallottini2016}). 
More specifically, \cite{Dekel2009} claim that about 30\% of the inflowing gas is converted into stars along the filamentary streams, before reaching the central massive galaxy.
In support of this scenario, \cite{Bouche2013,Bouche2016} report, through
absorption spectroscopy at high-z, the detection of inflowing gas significantly enriched in metals ($\rm Z\sim 0.4 \, Z_{\odot}$).
This scenario opens up new possibilities for the cosmic gas accretion to be investigated, within dense environment, by exploiting some of the classical tracers of star formation
and molecular gas. 

In this study we report ALMA observations of an extended molecular gas structure around
a massive star forming galaxy at $\rm z=3.47$ located in an overdense region (see Sec.~\ref{sec:data}), which  may provide new insights about gas accretion onto primeval galaxies.
\\\\
Throughout the paper, we assume a $\rm \Lambda$CDM cosmology with $\rm \Omega_m=0.3$, $\rm \Omega_{\Lambda}=0.7$ and $h=70 \,\rm km\,s^{-1}$. 
One arcsec at $\rm z\sim3.5$ corresponds to $\rm \sim 7.47 \,kpc$.
%%%%%%%%%%%%%%%%%%%%%%%%%%%%%%%%%%%%%%%%%%%%%

\section{Candels-5001: a massive galaxy in an overdense region at $ \lowercase{z}=3.47$}\label{sec:data}

Candels-5001 (ID 4417 in the GOODS-MUSIC catalogue; \citealp{Grazian2006}) is a Lyman-break-selected star forming galaxy at redshift $\rm z=3.47$ (measured through the [OIII]5007 $\rm \mathring{A}$ transition; \citealp{Maiolino2008}).  \newline
Candels-5001 is the most massive galaxy  ($\rm M_{\star} \sim 1.9 \times 10^{10}\,M_{\odot}$; \citealp{Santini2015}) in the redshift range $\rm 3<z<4$ within the whole  150 arcmin$^2$ covered by the Great Observatories Origins Deep Survey South field (GOODS-S).
It is actively forming stars at a rate of $\rm \sim 200 -250 \,M_{\odot} yr^{-1}$ (\citealp{Troncoso2014}).
This galaxy is also detected in the X-rays; however \cite{Fiore2012} exclude the presence of an AGN and show that the X-ray emission is fully consistent with what expected from the observed star formation rate.\newline
Table ~\ref{tab:candels_only} summarizes the main properties of the galaxy. \newline
Two star forming lower mass companions are located at a distance of about $\rm 7-15 \,kpc$ (i.e. $\sim1-2$ arcsec) from Candels-5001 and are likely in the process of merging (as suggested by their kinematics; see Sec.~\ref{sec:kinematics}).

\begin{figure}
	\centering
	\includegraphics[width=\columnwidth]{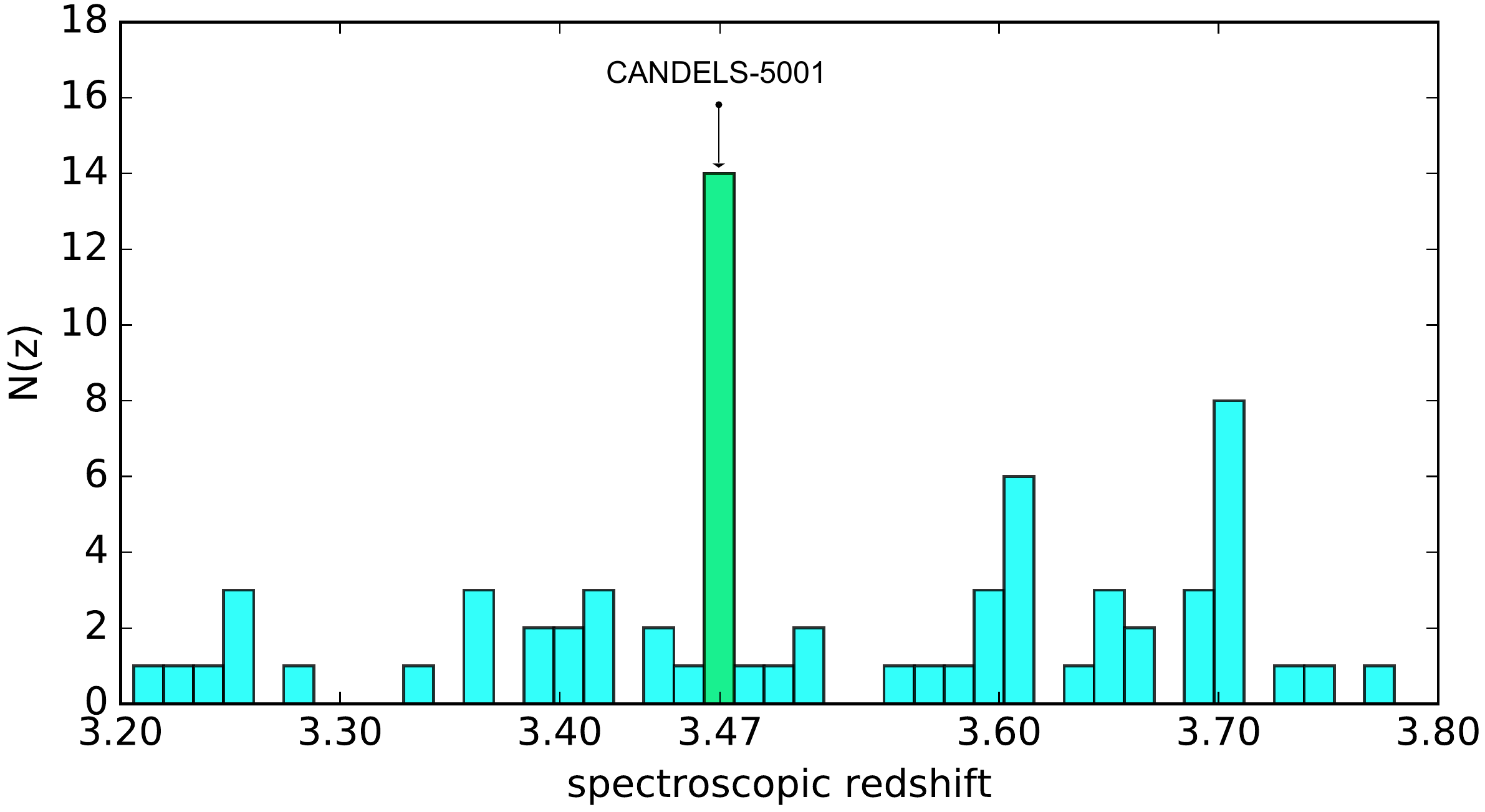}
	\caption{Distribution of spectroscopic redshifts of the galaxies optically identified in the GOODS-S field. The arrow indicates the redshift of Candels-5001.}
	\label{fig:overdensity_noCO}
\end{figure}

Candels-5001 is located in a large scale overdensity of galaxies traced by a clear, prominent spike in the distribution of spectroscopic redshifts over GOODS-S\footnote{We collected all available spectroscopic redshifts for galaxies around $\rm z\sim3-4$ in GOODS-S from the GOODS Multiwavelenght Southern Infrared Catalog (GOODS-MUSIC; \citealp{Santini2009}).} (Fig.~\ref{fig:overdensity_noCO}), suggesting it is the massive central galaxy in a forming protocluster. This has been recently confirmed by the analysis of \cite{Franck2016}, based on an accurate galaxy overdensity ($\rm \delta_{gal}$) selection criterion applied on a catalogue of spectroscopically identified candidate protoclusters.\newline
In such an overdense region, metal and dust enrichment in the IGM is supposed to be efficiently enhanced by galactic winds and outflows. \newline
This peculiar environment, combined with the achieved ALMA sensitivity and angular resolution ($\sim1''$), makes the selected target the right candidate to investigate cosmic gas accretion at high redshift by observing the molecular gas phase. 
%%%%%%%%%%%%%%%%%%%%%%%%%%%%%%%%%%%%%%%%%%%%%
%%%%%%%%%%%%%%%%%%%%%%%%%%%%%%%%%%%%%%%%%%%%%

%%%%%%%%%%%%%%%%%%%%%%%%%%%%%%%%%%%%%%%%%%%%%
\section{ALMA Observations and Data Reduction} \label{sec:observations}
%%%%%%%%%%%%%%%%%%%%%%%%%%%%%%%%%%%%%%%%%%%%%
\begin{figure}
	\centering
	\includegraphics[width=\columnwidth]{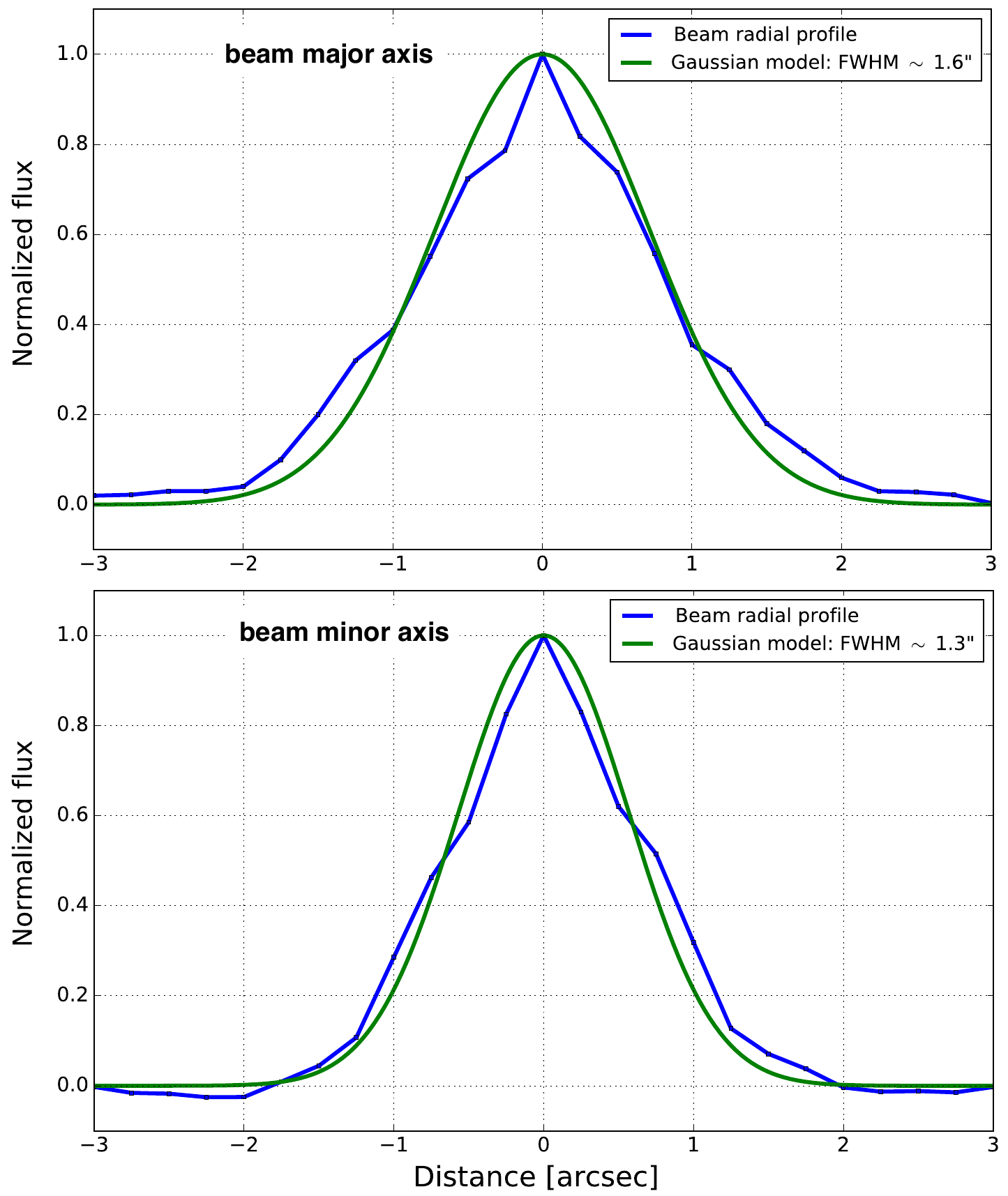}
	\caption{Upper (lower) panel: the beam radial profile along its major (minor) axis and a gaussian profile are shown. In both cases there is a deviation from gaussianity on small distances, allowing for some sensitivity on scales smaller than the beam FWHM.}
	\label{fig:beamProfile}
\end{figure}
%%%%%%%%%%%%%%%%%%%%%%%%%%%%%%%%%%%%%%%%%%%%%
%%%%%%%%%%%%%%%%%%%%%%%%%%%%%%%%%%%%%%%%%%%%%
\begin{figure*}
	\includegraphics[width=0.8\textwidth]{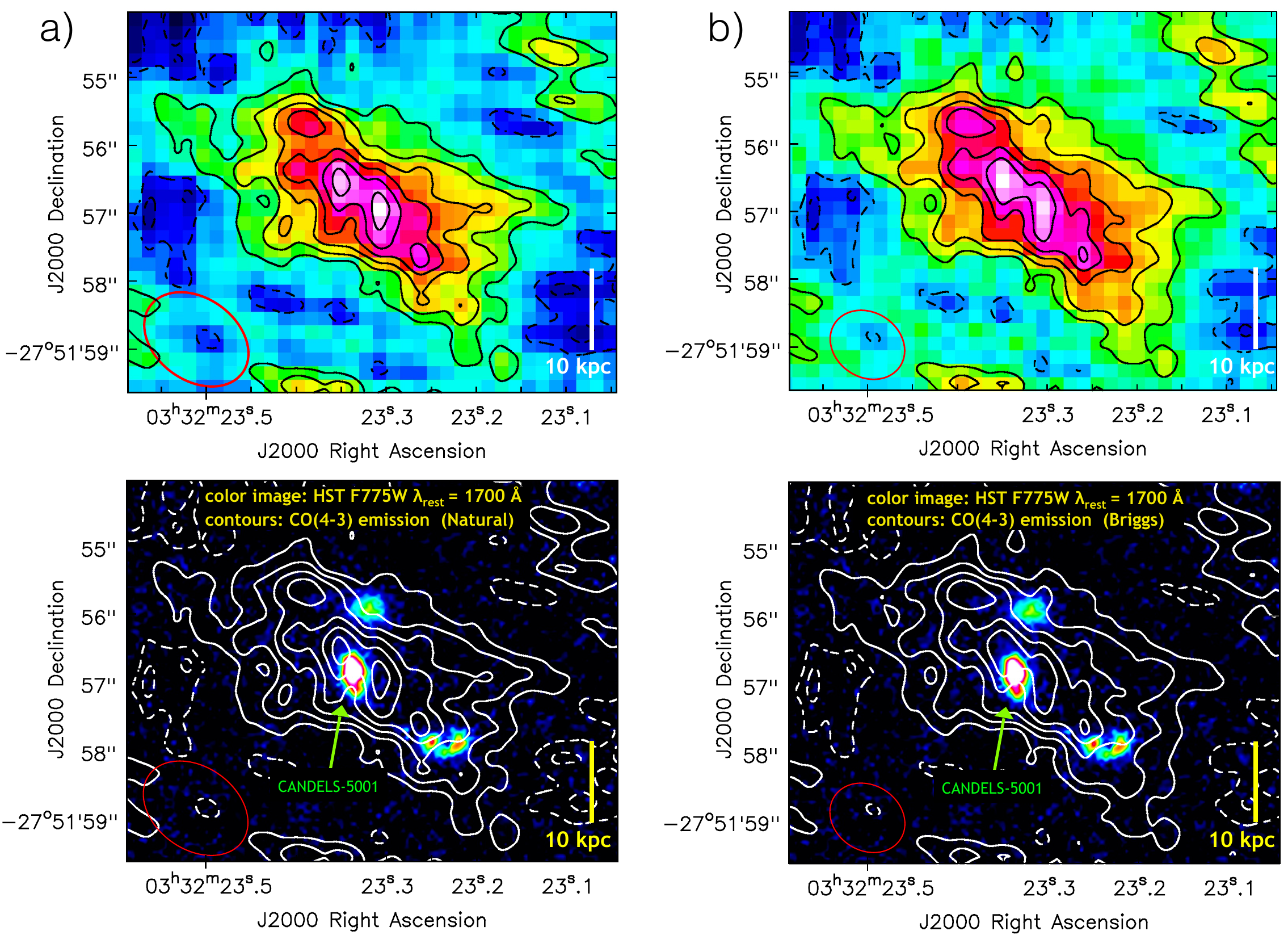}
	\caption{\textbf{a)} Upper panel: CO J=4-3 map of the region around Candels-5001 obtained through natural weighting  of the visibilities.
		The CO surface brightness contours range from $1\sigma$ to $6\sigma$, where
		$\rm \sigma = 22 \, mJy \, km \, s^{-1} beam^{-1}$ is the CO map rms. Dashed contours indicate negative fluxes.
		Lower panel: the same CO contours are shown overlaid onto the F775W HST image (color). \textbf{b)} Upper panel: CO J=4-3 map obtained through the Briggs weighting (robustness parameter: +2). Contours range from $1\sigma$ to $5\sigma$, where $\rm \sigma = 24.5 \, mJy \, km \, s^{-1} beam^{-1}$. In this image the extended emission is even more clearly resolved, over more than ten independent beams, although the sensitivity is somewhat lower. 
		Lower panel: same background colour image of a), with surface brightness contours of CO J=4-3 Briggs weighted map overlaid. The beam sizes  of the ALMA maps are given in the bottom-left corners.}
	\label{fig:COflux_maps}
\end{figure*}
%%%%%%%%%%%%%%%%%%%%%%%%%%%%%%%%%%%%%%%%%%%%%
Observations were carried out during ALMA cycle 1, in November 2013, making use of 30 antennas ($\rm 12\,m$ diameter), with a maximum baseline of $\rm 1.2\,km$, but most of the antennae were distributed in a relatively compact configuration. 
This results into a synthesized beam full-width at half-maximum (FWHM) of about 1.5$''$ in size, if using the natural weighting of the visibilities, but maps with a resolution of 1$''$ can be achieved by using different (Briggs, with robustness parameter of 2) weighting of the baselines (as discussed later on). 
The antennae were pointed at the optical position of Candels-5001. 
A total bandwidth of 7.5 GHz was employed in Band 3, using 4 wavebands between 90 GHz and 106 GHz. 
The primary beam FWHM at such frequencies is $\sim60''$. 
One of the wavebands was centred at 103.1 GHz, which is the frequency of the CO J=4-3 transition at the target redshift, i.e. z=3.473.
The other sidebands were used for continuum measurements.
A total of 116 minutes of on-source integration time, distributed over three execution blocks ($\sim 40$ minutes each), were obtained. \newline
The ALMA observatory staff performed initial data calibration, as part of standard data processing and delivery. 
The calibrated visibility data have been then re-analyzed, performing additional flagging of bad channels, using \texttt{CASA} version 4.5.3 (\citealp{Macmullin2007}).
The data have been imaged initially using the \texttt{CASA} natural weighting of the visibility data to create continuum map and datacubes. 
We achieve an rms sensitivity of  $\rm 9.6\, \mu Jy/beam$ for the continuum image and  $\rm 6 \,mJy/beam\,km\,s^{-1}$ in the waveband centred on the CO J=4-3 transition, where the latter rms is given in spectral bins of 20 $\rm km\,s^{-1}$ ($\sim7$ MHz). 
The sensitivity in the other three sidebands is lower since the noise is, on average, 10\% higher.  The absolute flux systematic uncertainty (associated with the flux calibration) is $\sim$10\%. \newline
A slight astrometric alignment of less than 0.5$''$ in the NW direction has been performed to match the ALMA data with the optical images (see Appendix~\ref{sec:appendix_offset} for a discussion). 
This level of systemic offsets between optical and ALMA data have been found in other cases \citep[e.g.][]{Dunlop2016,Maiolino2015} and are likely associated with either the astrometric accuracy of the ALMA phase calibrator and/or of the astrometric accuracy of the optical images.
However, we emphasize that the key results are not affected by the slight astrometric adjustment.
\newline
We note that the small relative number of antennae on long distances in our configuration results in a poorly covered visibilities plane on long baselines and, in turn, in a synthesized beam which is not Gaussian, as illustrated in Fig~\ref{fig:beamProfile}. These long baselines provide some sensitivities on scales smaller than the beam FWHM.
%%%%%%%%%%%%%%%%%%%%%%%%%%%%%%%%%%%%%%%%%%%%%
\section{Results}

\subsection{Molecular gas structure extending on 40 kpc scale around the central galaxy }\label{sec:extended}

\begin{table}
	\centering
	\caption{Summary of the properties of the extended molecular gas structure.}
	\label{tab:candels_40}
	\begin{tabular}{ccc} % four columns, alignment for each
		\hline
		Parameter &  Global Structure & Extended Component\\
		& $\rm 0<R\lesssim20 \,kpc$ & $\rm 2<R\lesssim20\, kpc$ \\
		\hline
		Line Flux CO(4-3)  & $0.30\pm 0.05$ & $0.17\pm 0.04$ \\
		Jy $\rm km\,s^{-1}$  &  & \\  [0.2cm]
		Molecular Gas Mass \ & $11.60^{+0.18}_{-0.29}$ & $11.49^{+0.19}_{-0.28}$\\
		log($\rm M_{H_2}/M_{\odot}$) &   & \\  [0.2cm]
		$\rm F_{cont}$(103~GHz)   & $75.4\pm 9.6$ & $39.3\pm 9.6$\\ 
		$\rm \mu Jy$ &   & \\  [0.2cm]
		Star Formation Rate & $198 \pm 63$ & $103 \pm 63$\\ 
		$\rm M_{\odot}\, yr^{-1}$  &   & \\  [0.1cm]
		\hline
	\end{tabular}
\end{table}
%%%%%%%%%%%%%%%%%%%%%%%%%%%%%%%%%%%%%%%%%%%%%
%%%%%%%%%%%%%%%%%%%%%%%%%%%%%%%%%%%%%%%%%%%%%
\begin{figure*}
	\includegraphics[width=0.8\textwidth]{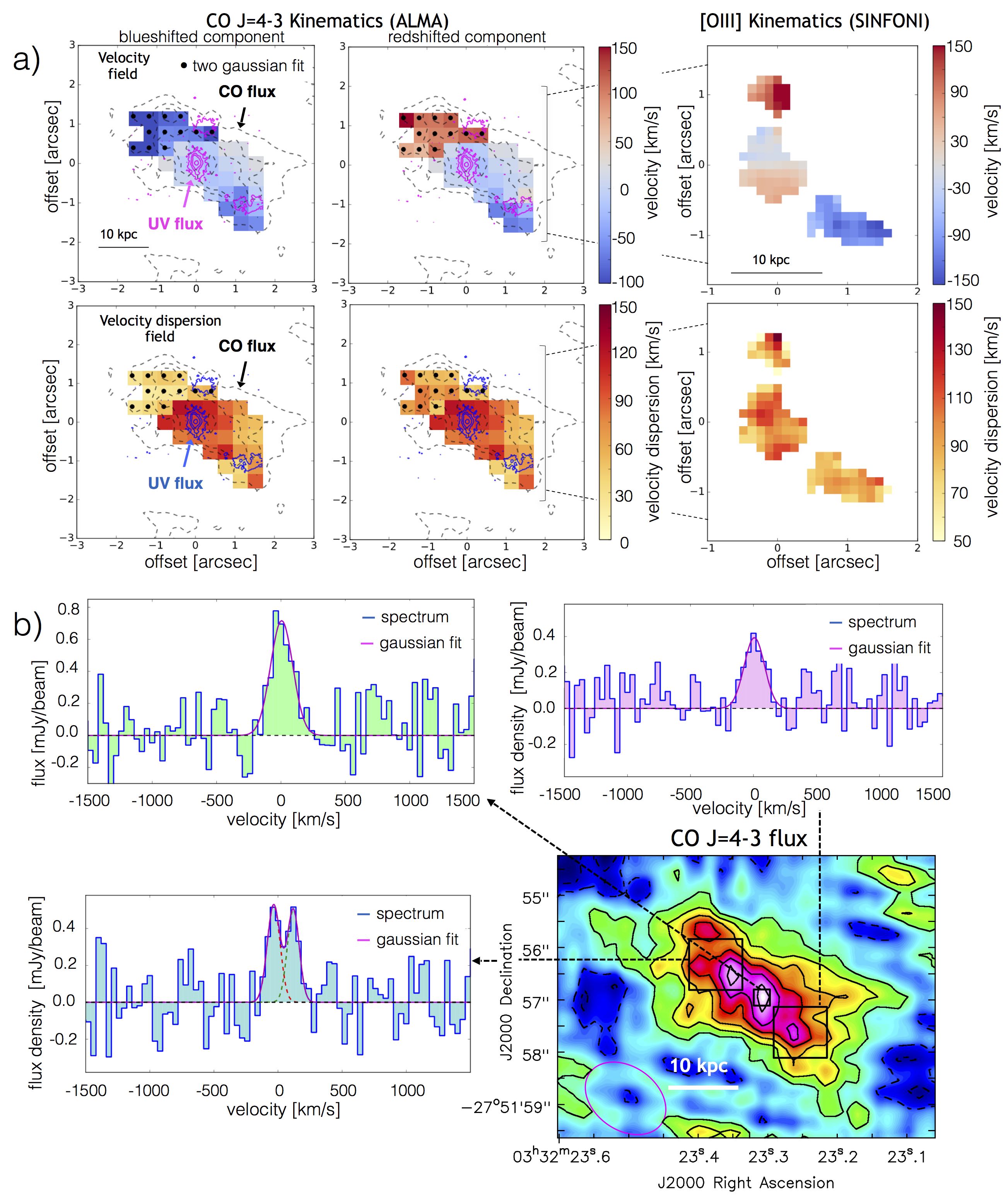}
	\caption{\textbf{a)} Left panels: CO J=4-3 velocity field and velocity dispersion. Dots indicate regions where a fit with two Gaussian components is required; in these cases
	the two kinematics components (\textquoteleft blueshifted\textquoteright\, and \textquoteleft redshifted\textquoteright) are shown separately. The solid contours show the rest-frame UV emission traced by HST. Right panels: velocity field and velocity dispersion of the three galaxies traced by the [OIII]5007$\rm \mathring{A}$ optical nebular emission line (\citealp{Maiolino2008}, \citealp{Troncoso2014}).
    \textbf{b)} Bottom right panel: CO J=4-3 surface brightness within the central few tens kpc around Candels-5001 (same levels as Fig.~\ref{fig:COflux_maps}a). Top/ bottom left panels: spectra extracted from the central region and from two outer regions, one of the latter showing indication of a double peaked profile.}
	\label{fig:kinematics}
\end{figure*}
%%%%%%%%%%%%%%%%%%%%%%%%%%%%%%%%%%%%%%%%%%%%%
A CO J=4-3 flux map was obtained by collapsing the channels containing line emission, ($\rm -140 < v < +140 \, km\,s^{-1}$; see Sec.~\ref{sec:kinematics}), using a datacube where the \texttt{CASA} natural weight of the visibilities was performed, resulting into the $\rm \sim 1.5''$ resolution mentioned in Sec.~\ref{sec:observations}  (specifically beam of $\rm 1.6'' \times 1.3''$ with PA=$54.32^\circ$).
The resulting CO flux map is shown in Fig.~\ref{fig:COflux_maps}a (left panel).
The CO emission is clearly extended over about 40 kpc in an elongated structure oriented NE-SW. 
In the bottom image of Fig.~\ref{fig:COflux_maps}a the same contours are overlaid onto an optical \textit{Hubble Space Telescope} (HST) image (F775W), tracing the rest-frame UV emission.  
We note that the CO map reveals some structures on scales smaller than the beam FWHM; this is a consequence of the presence of some long baselines that provide some sensitivity to small scales and make the beam profile non-Gaussian, as discussed in the previous section.
\newline
Fig.~\ref{fig:COflux_maps}b shows the same maps obtained with a \texttt{CASA} Briggs weight (i.e. weighting less the more compact baselines), which results into higher angular resolution (beam of 1$''$), although at expenses of a somewhat lower sensitivity. 
The latter image shows even more clearly that the extended emission is fully resolved, over more than ten independent beams.
The flux assigned to Candels-5001 has been estimated collecting the total flux within an ALMA synthesized beam centred on the rest-frame UV emission peak from HST images.   Interestingly, although the CO map peaks are associated with the massive star forming galaxy,  Candels-5001 accounts for only about 40\% of the CO emission in the central 40 kpc;  the remaining CO emission ($\sim 60\%$) is distributed in the CGM within the elongated structure, and it is not directly associated with either the central target or its two companions.

Within this context, we note that the extended structure cannot be ascribed to molecular gas hosted in the three merging galaxies and \textquoteleft artificially\textquoteright\, smeared on a larger scale by the ALMA beam. This is discussed more in detail in the Appendix~\ref{sec:appendixsmoothing} through a simple simulation.

We calculated the total molecular mass in the 40 kpc structure traced by the CO J=4-3 emission, conservatively assuming for such a system the same gas metallicity inferred for the central and nearby galaxies ($\rm \sim 0.5\, Z_{\odot}$, e.g. \citealp{Maiolino2008}, \citealp{Troncoso2014}, and validated by the Te-based calibration in \citealp{Curti2016}. This is a conservative assumption as the surrounding gas has probably lower metallicity, although \citealp{Bouche2016} has found evidence for inflowing gas with similar metallicity)
and using a metallicity dependent CO-to-H$_2$ conversion factor ($\rm \alpha_{CO}\sim10\pm2$; \citealp{Wolfire2010}; \citealp{Feldmann2011}; \citealp{Saintonge2013}; \citealp{Bolatto2013}). 
The inferred total mass of molecular gas within the 40 kpc wide structure is about $\rm 2-6 \times 10^{11} M_{\odot}$. 
The given range of gas masses reflects the uncertainty on $\rm \alpha_{CO}$ and on the CO J=4-3/CO J=1-0 lines flux ratio, for which we have assumed values ranging from that typically observed in normal (Milky Way-like) star forming galaxies to that observed in starburst and high-z submillimeter galaxies (\citealp{Carilli2013}). 
Table~\ref{tab:candels_40} contains detailed information on the extended elongated molecular gas structure surrounding Candels-5001. 

It is interesting to note that, as mentioned above, about 40\% of this molecular gas is associated with the central galaxy, implying that this is a very gas-rich
galaxy ($\rm M_{gas}/M_{\star}\sim 6$, Table~\ref{tab:candels_only}). This is not unprecedented among high redshift galaxies \citep[e.g.][]{Genzel2015} and it
reveals that this galaxy is in an early evolutionary phase.
%%%%%%%%%%%%%%%%%%%%%%%%%%%%%%%%%%%%%%%%%%%%%

\subsection{Kinematics properties of the extended gaseous structure} \label{sec:kinematics}
\begin{figure}
	\centering
	\includegraphics[width=0.9\columnwidth]{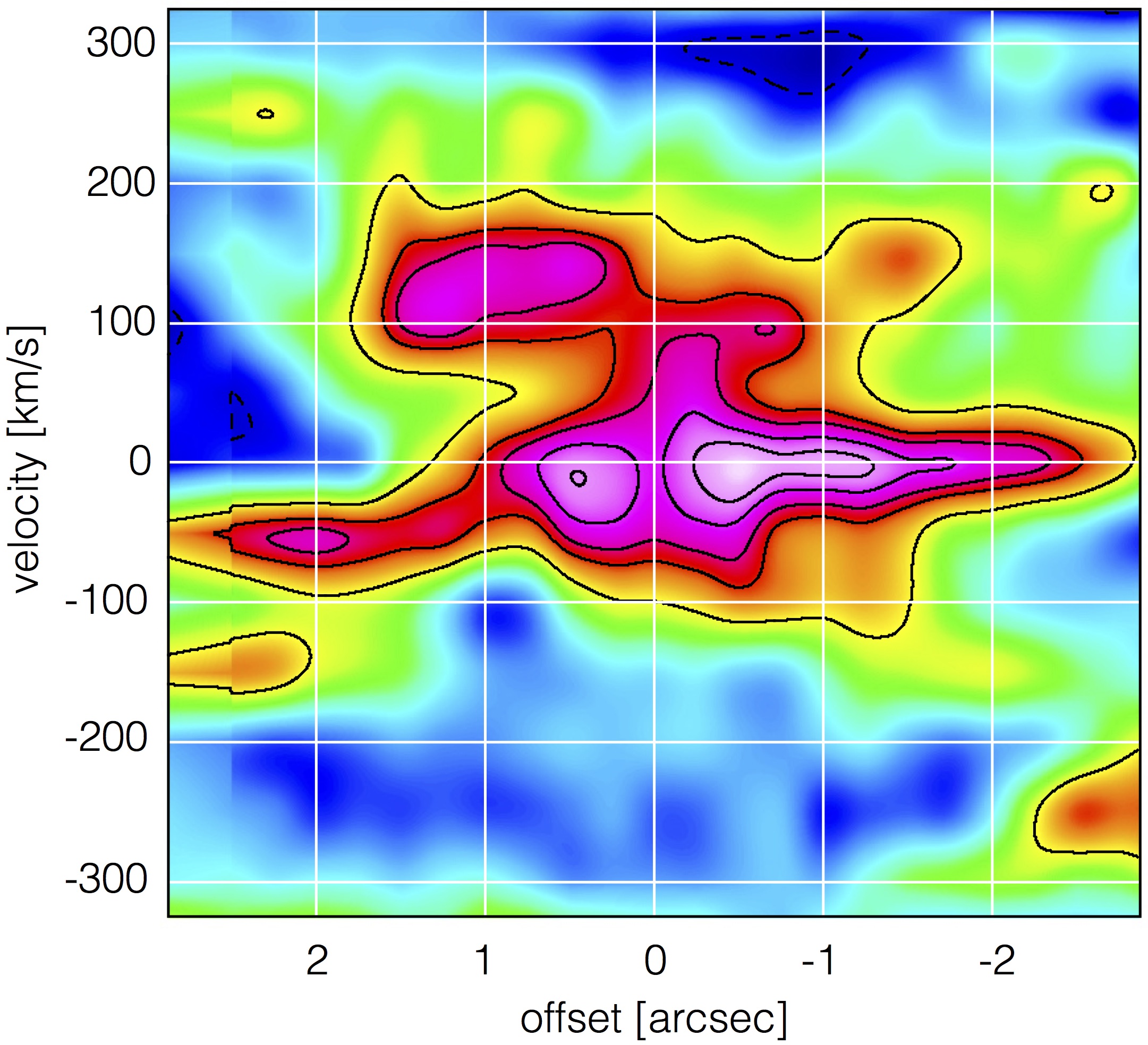}
	\caption{CO J=4-3 position-velocity diagram along the major axis of the extended structure. The diagram is inconsistent with simple coherent rotation and further showing sub-structures with different velocities at the same location.}
	\label{fig:pv_diagram}
\end{figure}
%%%%%%%%%%%%%%%%%%%%%%%%%%%%%%%%%%%%%%%%%%%%%
Velocity and velocity dispersion maps of the 40 kpc-scale gaseous structure were obtained by fitting one or two Gaussians to the line profiles and they are shown in the left panels of Fig.~\ref{fig:kinematics}a. 
While the overall dynamics broadly follows the velocity distribution of the three optical galaxies (Fig.~\ref{fig:kinematics}a, right panels), the external region North-West of Candels-5001 appears to be characterized by two different kinematic components. 
This is clear from the double peaked profile of the spectrum extracted from the area described above (see lower left panel of Fig.~\ref{fig:kinematics}b) and in the position-velocity (p-v) diagram along the major axis (Fig.~\ref{fig:pv_diagram}) of the 40 kpc gaseous structure.
The regions in which a double-Gaussian fit is required are marked with black dots in Fig.~\ref{fig:kinematics}a and the kinematics of the two components is shown separately in Fig.~\ref{fig:kinematics}a (left and central panels). \newline
The p-v diagram does not show any signature of coherent rotation around the massive galaxy. Instead  it clearly shows  sub-structures with different velocities at the same location. Overall these results highlight an irregular kinematics of the extended elongated molecular structure. \newline
We note that the  limited angular resolution of our ALMA data (with respect with the SINFONI observations of Candels-5001) does not allow us to resolve the internal kinematic structure of the molecular gas within the individual galaxies and, in particular, it does not resolve their internal rotation curves that are instead traced by their optical nebular lines (\citealp{Maiolino2008}) (Fig.~\ref{fig:kinematics}a, right panels).\newline
The measured CO line velocity dispersion appears to be low, ranging between 30 and 100 $\rm km\,s^{-1}$ over the entire region (Fig.~\ref{fig:kinematics}a, lower panels). 

%%%%%%%%%%%%%%%%%%%%%%%%%%%%%%%%%%%%%%%%%%%%%
\subsection{Continuum thermal emission within the 40~kpc extended structure} \label{sec:continuum}

%%%%%%%%%%%%%%%%%%%%%%%%%%%%%%%%%%%%%%%%%%%%%
\begin{figure}
	\centering
	\includegraphics[width=1\columnwidth]{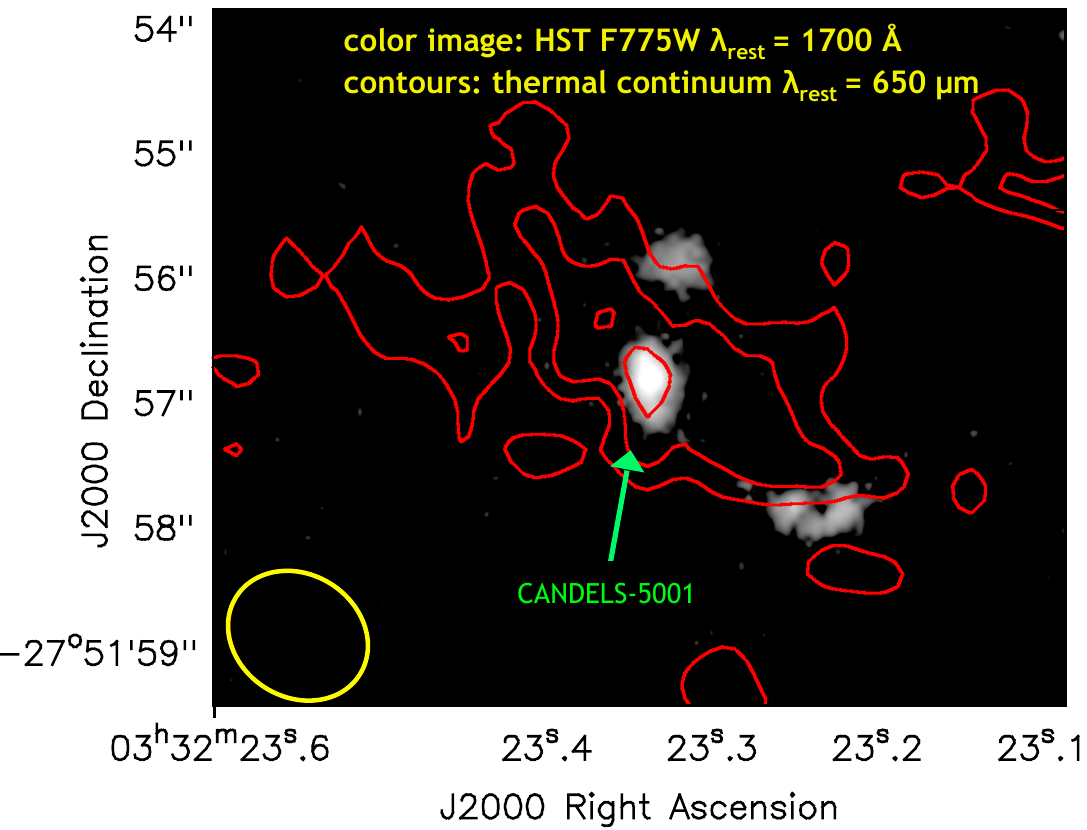}
	\caption{Continuum emission at 103 GHz, corresponding to $\rm \lambda_{rest} \sim 650\, \mu m$. Contours show the surface brightness at  1.5 $\rm \sigma$, 2.5 $\rm \sigma$ and 3.5 $\rm \sigma$, where $\rm \sigma = 9.6 \,\mu Jy beam^{-1}$ is the map rms. No negative fluxes at these levels are present in this area.}
	\label{fig:continuum}
\end{figure}
%%%%%%%%%%%%%%%%%%%%%%%%%%%%%%%%%%%%%%%%%%%%%
\begin{figure}
	\centering
	\includegraphics[width=1\columnwidth]{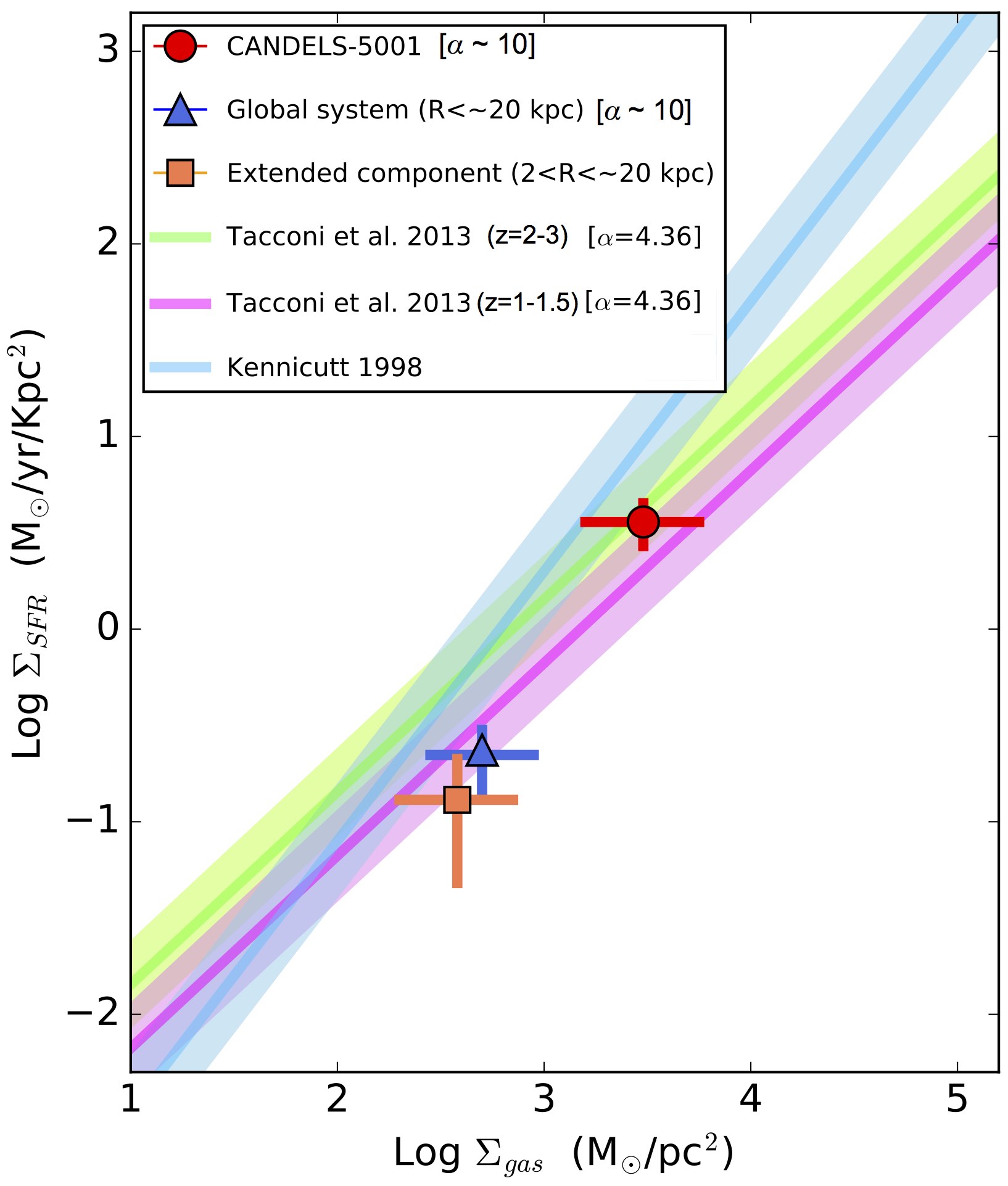}
	\caption{Location of the central massive galaxy and of the 40 kpc extended structure relative to the local and high-z Schmidt-Kennicutt relations between surface density of SFR and surface density of molecular gas.}
	\label{fig:KS1}
\end{figure}
%%%%%%%%%%%%%%%%%%%%%%%%%%%%%%%%%%%%%%%%%%%%%
We detect continuum thermal dust emission (at $\rm 103\,GHz$, corresponding to $\rm \lambda_{rest} \sim 650\, \mu m$) associated with the extended molecular gas structure traced by the CO emission.
This is shown in Fig.~\ref{fig:continuum} (with Briggs weighting), illustrating that continuum emission is associated with the central massive galaxy, but also clearly extended along the surrounding structure (without being associated with any of the two companion galaxies). \newline
In other works thermal emission at these wavelengths has been used  as a tracer of star formation (whose UV radiation is responsible for heating the dust) and/or as tracer of the dust mass.  
\cite{Scoville2014, Scoville2016} have shown that, for most galaxies, in the Rayleigh-Jeans region a constant dust temperature of 25~K can be assumed and, therefore, the continuum emission can be used as a tracer of dust mass.
However, this assumptions may break down in an extended primordial structure as the one observed by us on 40~kpc scale, both because of the low metallicity and because the ISM may have physical conditions different than in normal galaxies.

We have combined the ALMA continuum detection with \textit{Herschel} data of the same galaxy. 
The \textit{Herschel} beam is so large that it is not possible to disentangle the
emission on 40~kpc scales from the emission associated with
central galaxy Candles-5001.
We have therefore performed a fit of the IR-to-millimeter SED of the entire 40~kpc emitting region by using the fitting code CIGALE (\citealp{Burgarella2005}) to retrieve the main parameters of the system. 
We obtain that the inferred total star formation rate is $\rm 198\pm 63 ~M_{\odot}yr^{-1}$.
This is roughly consistent, although somewhat lower, to the sum of the star
formation rates inferred from the optical-UV SED fitting or the
(extinction-corrected) H$\beta$ emission of the three merging galaxies, which is
about $\rm 325\pm 120 ~M_{\odot}yr^{-1}$ (\citealp{Troncoso2014}).
The slightly lower SFR inferred from the far-IR emission can be explained in terms of lower dust content which absorbs less UV radiation than in metal rich galaxies.
From the same fit we infer a dust mass of $\rm M_{dust}=1.58~(\pm 0.57) ~\times 10^9~M_{\odot}$. 
Assuming a metallicity of half solar (as inferred for Candles-5001) and the
metallicity dependent H$_2$-to-dust ratio derived by \cite{Leroy2011}, we infer a
total molecular mass in the central 40~kpc of $\rm 2.1~(\pm 0.75)~\times
10^{11}~M_{\odot}$, which is roughly consistent with what inferred through the CO
emission, within uncertainties (Table \ref{tab:candels_40}).  \newline
We cannot perform a similar SED analysis on spatially resolved basis due to the low angular resolution of the \textit{Herschel} data. 
However, to first order, we can assume the same SED fitted to the total IR-to-mm
emission and split it into a component associated with Candles-5001 and an other one associated with the extended emission ($\rm 2<R<20~kpc$), with the proportions inferred from the ALMA continuum flux distribution.
By using this approach, we infer that the the SFR in the extended component is $\rm \sim 100\,M_{\odot} yr^{-1}$.
This is similar to the SFR inferred for the central galaxy, but distributed on much larger scales, implying an average surface density of SFR of $\rm 0.1 \,
M_{\odot}~yr^{-1}\, kpc^{-2}$.
Such extended star formation is not observed in the HST rest-frame UV image. 
From the latter data we infer an upper limit of $\rm 0.03 \, M_{\odot} yr^{-1}\, kpc^{-2}$. 
Such a discrepancy may be ascribed to dust extinction: assuming the dust attenuation curve typical of star forming galaxies (\citealp{Calzetti2000}), matching the two observations requires a dust reddening of only $\rm E_{B-V}=0.1$, which is lower than that inferred for the optical star forming galaxies in the same region (\citealp{Troncoso2014}). 
The tentative detection of some weak diffuse emission in longer wavelength HST images (F160W, tracing continuum stellar emission at $\rm \lambda_{rest}\sim3400\rm \mathring{A}$; see Appendix ~\ref{sec:HST-Spitzer maps}) supports such dust-reddening scenario. 
Alternatively, dust in the extended structure may be significantly cooler than what observed in typical star forming galaxies, which would imply a significantly lower SFR. 

It is interesting to note that, with the $\rm \Sigma _{SFR}$ inferred above, the 40 kpc extended structure surrounding Candels-5001 is located slightly below (although still marginally consistent with) the Schmidt-Kennicutt relation between SFR and molecular gas (Fig.~\ref{fig:KS1}), suggesting that the diffuse star formation in such extended structure is somewhat less efficient than in normal galaxies (\citealp{Tacconi2013}; \citealp{Sargent2014}). 
If a fraction, or most of the continuum emission in the extended structure is not due to in-situ star formation, but heated by external optical-UV radiation field, then the inferred star formation efficiency would be even lower.

%%%%%%%%%%%%%%%%%%%%%%%%%%%%%%%%%%%%%%%%%%%%%
\subsection{Detection of CO systems on 550 kpc scale}\label{sec:COlargescale}

%%%%%%%%%%%%%%%%%%%%%%%%%%%%%%%%%%%%%%%%%%%%%
\begin{figure}
	\centering
	\includegraphics[width=\columnwidth]{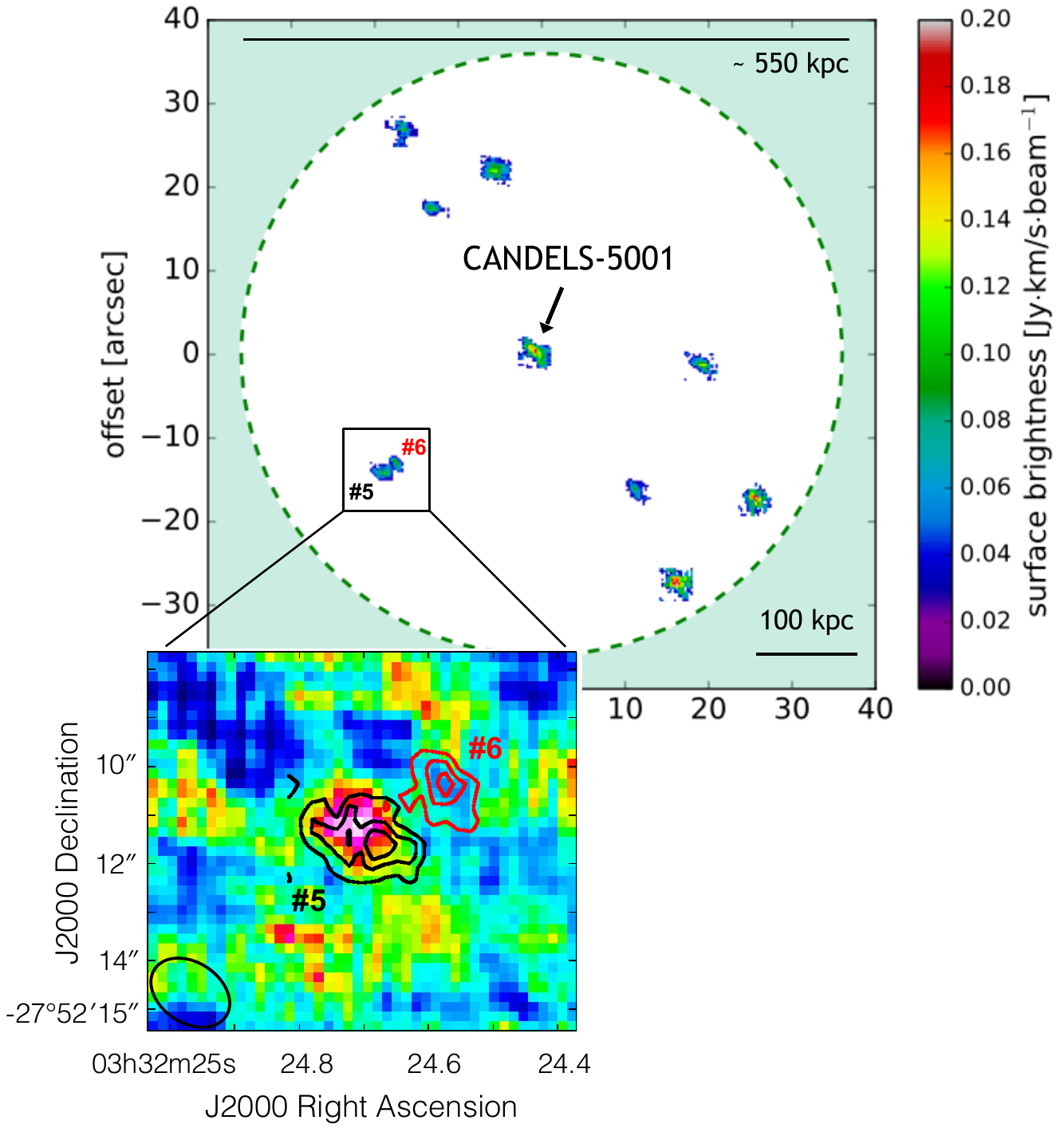}
	\caption{Distribution of CO emitting systems detected on the 550 kpc scale around Candels-5001.
	The dashed circle encloses the region where the primary beam response function is above 30\%. The image of each object is shown down to the 1$\rm \sigma$ level of the CO surface brightness. 
	The inset shows (in colour) the ALMA continuum emission map for the galaxy (in addition to Candels-5001 and its surrounding 40 kpc) clearly detected in continuum. Black and red contours show CO emission systems \#5 and \#6 (see Appendix~\ref{sec:largescale}), respectively, at levels of 2 $\rm \sigma$, 3.5 $\rm \sigma$ and 5 $\rm \sigma$.}
	\label{fig:largescale1}
\end{figure}
%%%%%%%%%%%%%%%%%%%%%%%%%%%%%%%%%%%%%%%%%%%%%
%%%%%%%%%%%%%%%%%%%%%%%%%%%%%%%%%%%%%%%%%%%%%
\begin{figure}
	\centering
	\includegraphics[width=\columnwidth]{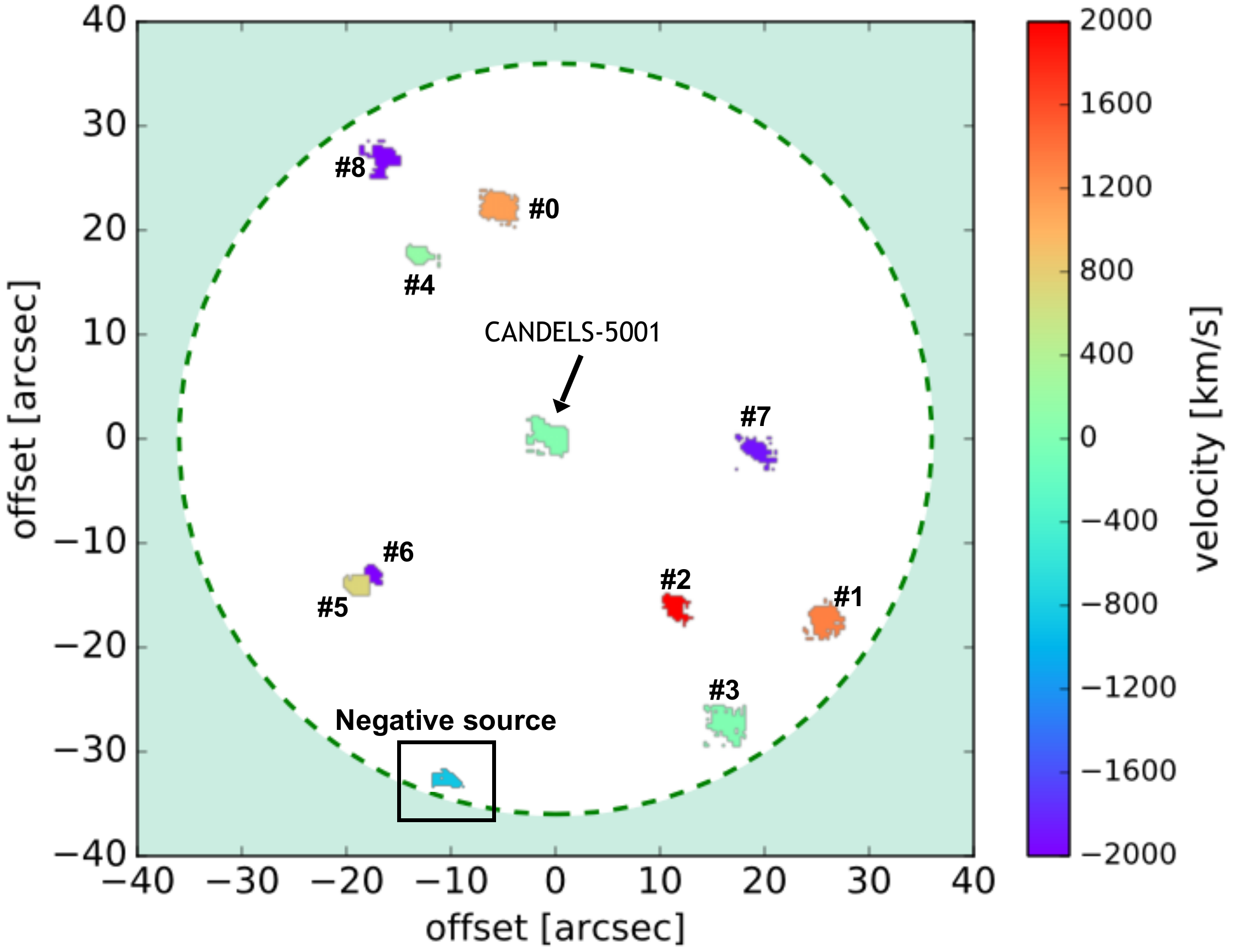}
	\caption{CO emitting systems colour-coded according to their velocity relative to Candels-5001. \textit{Black box}: the only negative source (i.e. detected in the inverted flux cube and with the same criteria used to detect the positive sources), in the waveband centred on the CO J=4-3 transition.}
	\label{fig:largescale2}
\end{figure}
%%%%%%%%%%%%%%%%%%%%%%%%%%%%%%%%%%%%%%%%%%%%%
%%%%%%%%%%%%%%%%%%%%%%%%%%%%%%%%%%%%%%%%%%%%%
\begin{figure*}
	\centering
	\includegraphics[width=0.9\textwidth]{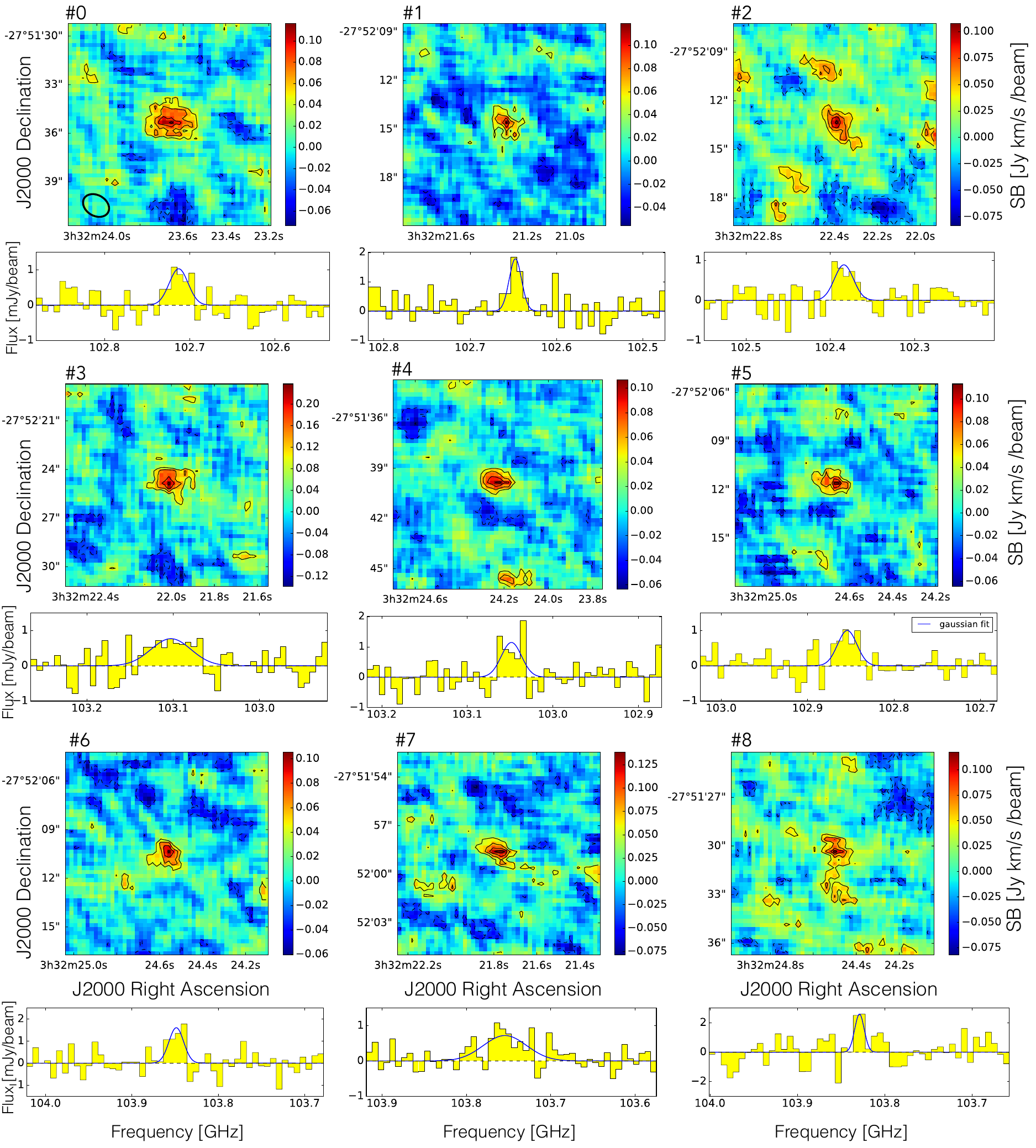}
	\caption{CO J=4-3 maps of the individual systems detected on large scales, along with their spectra. CO surface brightness contours are shown, at levels of $2\sigma$, $3\sigma$, $4\sigma$, $5\sigma$, where $\rm \sigma$ is the rms in each CO map. Dashed contours indicate negative fluxes. The beam size  of the  maps is given in the bottom-left corner of the upper-left panel.}
	\label{fig:COsystems1}
\end{figure*}
%%%%%%%%%%%%%%%%%%%%%%%%%%%%%%%%%%%%%%%%%%%%%
%%%%%%%%%%%%%%%%%%%%%%%%%%%%%%%%%%%%%%%%%%%%%
\begin{figure}
	\centering
	\includegraphics[width=1\columnwidth]{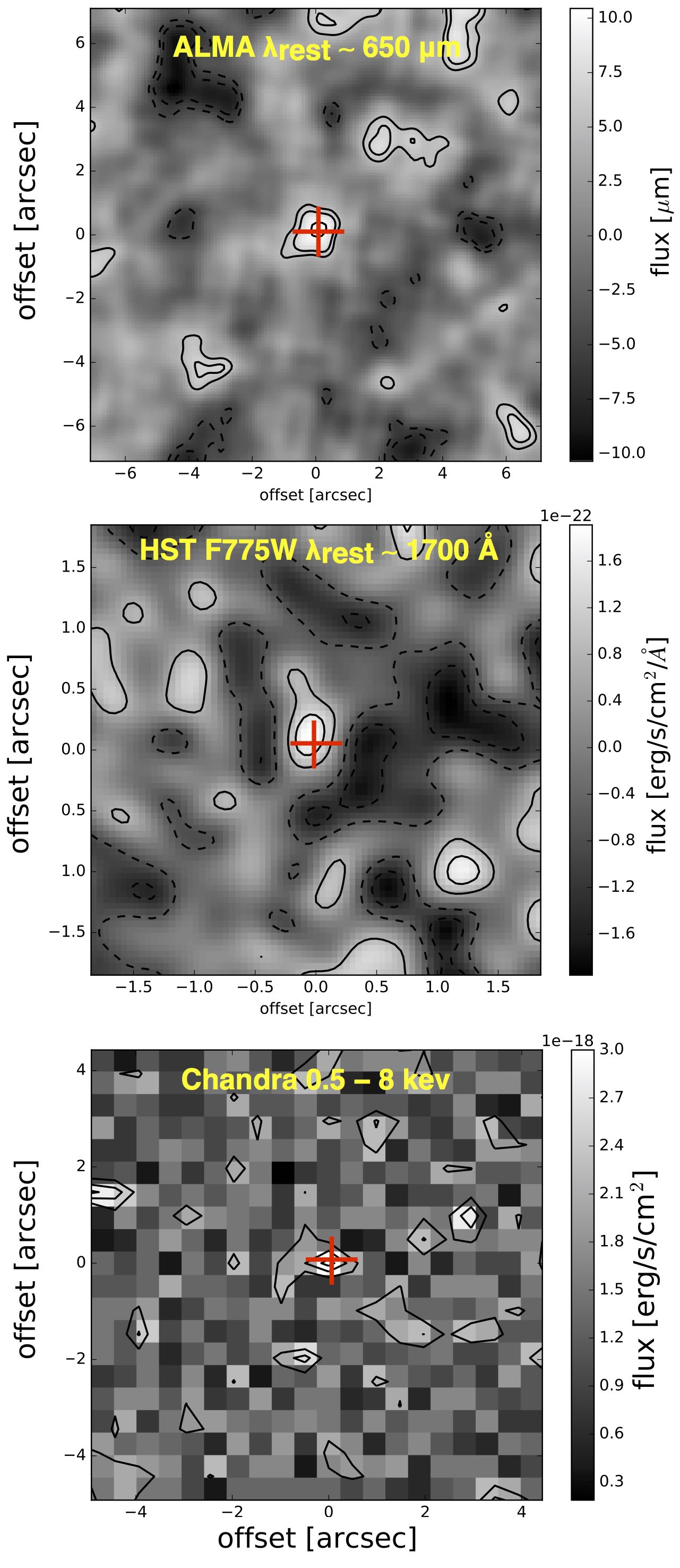}
	\caption{Upper panel: ALMA continuum stacked image of the large scale CO emitters (excluding system \#5, which is individually detected). Contours levels: $1\sigma$, $2\sigma$, $3\sigma$.
	Central panel: HST F775W stacked image of all CO sources on large scales. Contours levels: $1\sigma$, $2\sigma$. Lower panel: Chandra/ACIS X-ray stacked image of all CO sources. Contours levels: $1\sigma$, $2\sigma$, $3\sigma$.}
	\label{fig:stack}
\end{figure}
%%%%%%%%%%%%%%%%%%%%%%%%%%%%%%%%%%%%%%%%%%%%%
We have searched for CO emitting systems within the ALMA field of view by collapsing spectral channels of the natural weighted datacube for a resulting channel width ranging from $\rm 40\,km\,s^{-1}$ to $\rm 200\,km\,s^{-1}$ (although almost all detections are found with widths larger than $\rm 100\,km\,s^{-1}$) with the centre of the rebinned spectrum spanning the whole waveband, and searching for emission within the area where the primary beam response function is above 30\%, i.e. a diameter of $\sim$73$''$, corresponding to 550 kpc. 
Fig.~\ref{fig:largescale1} shows the distribution of the CO detections within the $\sim$73$''$ diameter (traced by the dashed circle) where the colour-coding of the detections gives the surface brightness.
%%%%%%%%%%%%%%%%%%%%%%%%%%%%%%%%%%%%%%%%%%%%%
The significance of each system was assessed by comparing it with the noise in the map extracted within the same spectral range. 
To select the reliable CO detections we used a threshold of $\rm 5.2\,\sigma$ for CO peak emission, with the justification that with this threshold the number of negative detections (i.e. number of sources detected with the same method in the flux-inverted cube) is less than 10\% of the positive ones. 
Indeed, following these criteria, we obtain 10 positive detections and only one negative detection within the spectral window centred onto the CO J=4-3 transition.
Moreover, such a negative detection is found at the very edge of the field, as shown in the black box inset of Fig.~\ref{fig:largescale2}.  \newline
This test indicates that the vast majority of the CO systems are not noise fluctuations and that the contamination from spurious detections is only about 10\% or less. \newline
Within the other three wavebands, the ratio between positive and negative detections tends to be close to one, as expected from typical Gaussian noise fluctuations. 
In such other spectral windows, as a consequence of the higher noise (relative to the band centred onto the CO J=4-3 transition), the number of positive and negative detections is about two or three per spectral band (by using the same \textquoteleft detection criteria\textquoteright\, discussed above for CO J=4-3). 
Such \textquoteleft detections\textquoteright\ in the other spectral bands (most of which must be associated with noise fluctuations) are isotropically distributed within the field of view. \newline
We note that the field of view does not contain any strong source that may produce prominent side-lobes (especially in the narrow bands used for the CO detections).

The zoomed map of each of the CO systems detected on large scales is shown in Fig.~\ref{fig:COsystems1} along with each individually extracted spectrum.

\subsubsection{Physical properties of the $\sim550\,kpc$-scale CO systems}

From Fig.~\ref{fig:largescale1} it is clear that the CO systems detected on 550~kpc scales are predominately distributed in the same NE-SW direction as the central  $\rm \sim$40 kpc extended molecular gas structure (Fig.~\ref{fig:COflux_maps}), strongly suggesting a physical link between them. 
Assuming metallicities similar to Candels-5001, and using the same calibrators discussed before,  we infer for these systems gas masses of about $\rm 10^{10} - 10^{11}\, M_{\odot}$. \newline
We note that the CO J=4-3 transition is highly sensitive to the excitation state of the emitting gas, and therefore, since these gas rich systems may be partially shock-heated (\citealp{Nelson2016}), this may result into an overestimation of their molecular mass content. \newline
In Appendix~\ref{sec:largescale} we show the thumbnails of the optical and near-IR HST/Spitzer images of these systems and their properties are also tabulated. 
Some of the systems show a potential HST and/or Spitzer counterpart (e.g. systems \#3 and \#4), especially if one allows for some offset between ALMA and optical/IR counterparts, which seem to be common among these sources \citep{Dunlop2016}.
ALMA continuum emission is clearly detected in system \#5 (illustrated in the zoomed inset of Fig.~\ref{fig:largescale1}), and marginal detections (at the 2.5$\sigma$ level) are also seen for systems \#0 and \#2.

The other systems are not  clearly detected individually, however the HST stacked image of all CO emitting sources, shown in Fig.~\ref{fig:stack} does show a 2.5$\sigma$ detection. 
The stacked ALMA continuum emission (excluding system \#5, which is individually detected) results into a $3\sigma$ detection
(Fig.~\ref{fig:stack}). Unfortunately, given the lack of information at other IR wavelengths, we cannot directly translate this
flux ($\rm \sim 11\, \mu Jy$) into SFR or dust mass. Yet, if we assume the same SED as the central region, then the implied average SFR of these
systems would be about $\rm30\, M_{\odot}~yr^{-1}$.
Interestingly, a $3\sigma$ detection is also obtained by the stacked \textit{Chandra}/ACIS X-ray image ($\rm 0.5-8~keV$), as
shown in Fig.~\ref{fig:stack}.
The stacked flux is consistent (\citealp{Ranalli2003}) with the X-ray emission expected from galaxies with an average SFR in the range of 50--120 $\rm M_{\odot}yr^{-1}$ (depending on whether the soft X-ray relation or the hard X-ray relation of \citealp{Ranalli2003} is used and also depending on the assumed X-ray slope). 
It is interesting to note that such SFR, combined with the average molecular gas mass inferred for these systems ($\rm \sim
1.5\times 10^{11}~M_{\odot}$), would imply that these systems have an average depletion time of 2~Gyr, i.e. are less efficient in
forming stars than normal (main sequence) galaxies at the redshift of the source \citep{Sargent2014}. If some of the X-ray flux is contributed
by obscured AGNs and/or if we take the SFR inferred from the stacked thermal continuum emission discussed above, then their depletion
timescale would be even longer.

The velocity map of the CO systems on large scale, relative to Candels-5001 and its surrounding molecular gas (Fig.~\ref{fig:largescale2}), shows that they uniformly cover the ALMA waveband, ranging velocities from few hundreds $\rm km \,s^{-1}$ up to almost $\rm \pm 2000\, km\,s^{-1}$. \newline
Some of the high velocity systems may be affected by the local environment and their velocity could be boosted as a consequence of close gravitational interactions.
This is clearly the case for the low mass, high velocity CO system \#6, located a few arcsecs from the more massive galaxy \#5.
The former is probably a satellite galaxy in the process of merging with the more massive nearby galaxy companion, and its peculiar high velocity may simply be the result of the local merging process.  \newline
We note that the high velocity spread of the large scale CO systems is comparable with the velocity dispersions observed in other high-z protoclusters (\citealp{Casey2015}; \citealp{Chiang2015}), and indeed it is consistent with the redshift range ($\rm \Delta z \sim0.015$) of the peak in the redshift distribution of the spectroscopically identified galaxies in the GOODS-S field (Fig.~\ref{fig:overdensity_noCO}). 

\begin{figure}
	\centering
	\includegraphics[width=\columnwidth]{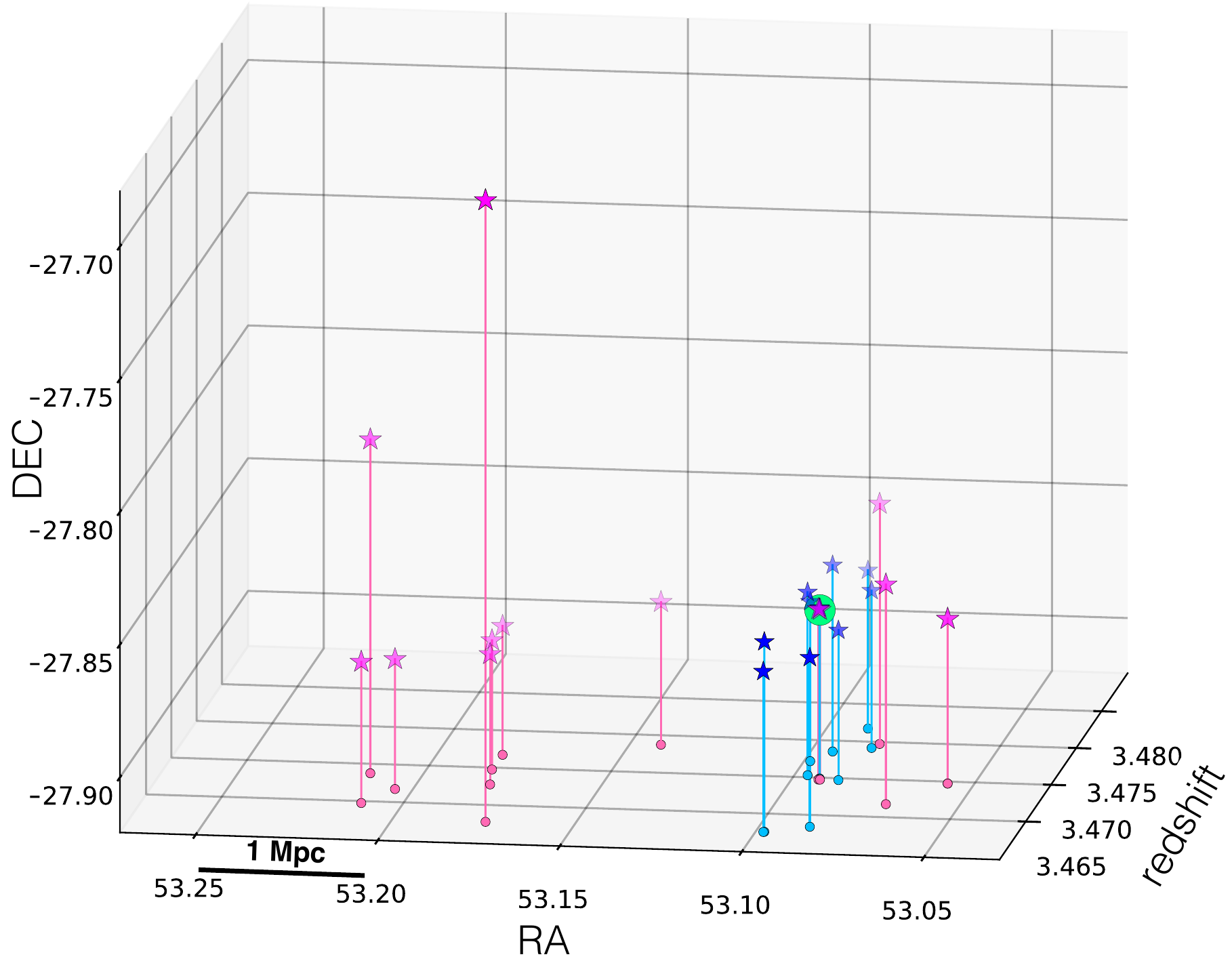}
	\caption{Distribution of the galaxies spectroscopically identified in the redshift spike    shown in Fig.~\ref{fig:overdensity_noCO} (magenta symbols), along with
	the distribution of the CO systems (blue symbols) in the Right Ascension - Declination - redshift diagram. The location of Candels-5001 is indicated with a green circle.}
	\label{fig:3Doverdensity}
\end{figure}

Therefore, the high velocity CO emitters may be actually tracing the gravitational field of the forming proto-cluster.
Fig.~\ref{fig:3Doverdensity} shows the three-dimensional distribution of the CO systems identified in the ALMA field of view (blue symbols), along with the distribution of the spectroscopically identified optical galaxies (magenta symbols) in GOODS-S, further confirming that all those objects in this narrow redshift interval belong to a proto-cluster, traced by an overdense region spanning a few Mpc in size.
%%%%%%%%%%%%%%%%%%%%%%%%%%%%%%%%%%%%%%%%%%%%%
%%%%%%%%%%%%%%%%%%%%%%%%%%%%%%%%%%%%%%%%%%%%%
%%%%%%%%%%%%%%%%%%%%%%%%%%%%%%%%%%%%%%%%%%%%%

\section{Discussion}\label{sec:discussion}

% disc scenario
%%%%%%%%%%%%%%%%%%%%%%%%%%%%%%%%%%%%%%%%%%%%%
Previous observations of CO rotational transitions at high-z have found molecular gaseous structures, in some cases quite extended, on scales of even about 10 kpc \citep{Tacconi2013,Hodge2012,Decarli2016}. 
Such systems have generally been interpreted as large rotating gaseous discs, in which
optical clumps are identified as regions of star formation occurring within such discs (\citealp{Genzel2011}).
Relatively large discs have also been identified in the ionized gas component, through $\rm H\alpha$ imaging (\citealp{Genzel2006, Genzel2010, Genzel2014}); the latter is mostly a proxy of star formation and probes only a very small fraction (the warm ionized phase) of the gas content in galaxies. \newline
Differently from these objects, our observations of the extended ($\rm \sim$40 kpc) elongated molecular gas structure surrounding Candels-5001 cannot be interpreted in terms of a single large gaseous rotating galaxy disc, in which the optically identified galaxies are sub-clumps. 
In the follow we describe the main inconsistencies with such a scenario. 
First, we note that molecular galaxy discs as large as 40 kpc have never been observed and neither expected by any theory or cosmological simulation, especially at $\rm z>3$ (the largest putative molecular disc found so far has a size of $\rm \sim 20\,kpc$, reported by \citealp{Tacconi2013}, and it is at $\rm z\sim1.2$, i.e. at a much later cosmic epoch). 
Even more importantly, as traced by the [OIII]5007$\rm \mathring{A}$ optical nebular line (SINFONI observations; Fig.~\ref{fig:kinematics}a), the individual optical systems (Candels-5001 and its two companions) have individually resolved rotation curves, which are inconsistent with the interpretation of them being sub-clumps of a single very massive galaxy disc.  
Finally, and most importantly, as already shown in Sec~\ref{sec:kinematics}  the CO position-velocity diagram along the major axis (Fig.~\ref{fig:pv_diagram}) is completely inconsistent with a single large coherently rotating disc. 

% stripped gas scenario
%%%%%%%%%%%%%%%%%%%%%%%%%%%%%%%%%%%%%%%%%%%%%
An alternative scenario in which the observed extended CO emission is tracing tidally stripped molecular gas is also not plausible. 
It would imply that the fraction of molecular gas mass stripped out of the galaxy by gravitational interactions with the two satellites would exceed 60\%. 
In contrast to the atomic gas phase (which is located primarily in the galaxy outskirts and more loosely bound), the molecular gas phase in galaxies is more centrally concentrated and more gravitationally bound.
Indeed, both observations and simulations show that, in merging/interacting galaxies,
only a tiny fraction, if any, of the molecular gas goes in tidal tails, while the bulk of the molecular gas becomes even
more centrally concentrated, as a consequence of angular momentum losses due to cloud-cloud collisions (\citealp{Ueda2014}; \citealp{Narayanan2006}; \citealp{Sanders1996}). \newline
% outflowing gas scenario
%%%%%%%%%%%%%%%%%%%%%%%%%%%%%%%%%%%%%%%%%%%%%

We also discard the possibility of the massive gas reservoir surrounding Candels-5001 being the product of starburst or stellar-feedback driven outflows.
The huge amount of mass found in the gaseous structure around Candels-5001 ($\rm M_{gas} \sim 1.5-4.5 \times 10^{11} M_{\odot}$, see Table ~\ref{tab:candels_40}) and the low CO line velocity dispersion ($\rm \lesssim 100 \,km\,s^{-1}$; lower panels of Fig.~\ref{fig:kinematics}a) are indeed inconsistent with systems where such phenomena really occur,  indicating that the extended structure is not associated with a molecular outflow.
Within this context it is also important to note that, as mentioned in Sec.~\ref{sec:data}, that \citealp{Fiore2012} excluded any possible AGN activity in Candles-5001, hence further discarding the scenario of a powerful AGN-driven outflow.

\subsection{Molecular gas in accreting streams}
% our interpretation
%%%%%%%%%%%%%%%%%%%%%%%%%%%%%%%%%%%%%%%%%%%%%
We suggest that our ALMA observations of the extended ($\sim$$\rm40\,kpc$) elongated gaseous structure surrounding Candels-5001, combined with its kinematic properties and the overdense environment, are consistent with the system tracing the inner and densest parts of large scale filamentary streams, feeding the central massive galaxy. \newline
The different kinematic components within such extended structure, identified in both the velocity field (Fig.~\ref{fig:kinematics}a) and the pv-diagram (Fig.~\ref{fig:pv_diagram}), indicate that Candels-5001 is likely at the convergence point of three intersecting filaments, as models predict for early massive galaxies (\citealp{Dekel2009}; \citealp{SanchezAlmeida2014}; \citealp{Keres2005}; \citealp{Nelson2016}; \citealp{Shen2013}; \citealp{vandeVoort2011}).\newline
If this interpretation is correct, then this would be one of the very first evidences of molecular gas accreting onto a massive galaxy in the early Universe, as well as the first detection of the gas reservoirs supposed to surround high redshift objects in terms of their molecular phase, which traces the most massive component.  \newline 
MRC 1138-262, often referred as the \textquoteleft Spiderweb Galaxy\textquoteright\, (\citealp{Pentericci2002}; \citealp{Miley2006}; \citealp{Venemans2007}; \citealp{Nesvadba2006}), a radio galaxy at the centre of a protocluster, could be another system in which accretion of cold gas may be occurring, although at a significantly lower redshift (z=2.16) than Candles-5001. 
However, observational evidence for cold accretion in this system is still unclear or marginal.
Indeed, observations of molecular gas of this system have either been at very low angular resolution ($\rm 9.5\times5.3$ arcsec = $\rm 80\times45$ kpc; \citealp{Emonts2013}), preventing the association of the molecular emission with intracluster molecular gas or with the blending of CO emission from the various merging galaxies optically identified within the beam of the radio observation, or have resulted in the detection of a few individual clumps associated with either AGN nuclei or clumps excited by the radio jets produced by the central radio galaxy (\citealp{Gullberg2016}). \newline
Our detection is likely resulting from a combination of several factors, specifically: 1) sensitivity enabled by ALMA; 2) instrument configuration (enabling the detection of diffuse/extended emission on large scales); 3) properly selected target at the proper cosmic epoch (a massive galaxy at the centre of a protocluster, at redshift z=3.47, when accretion from cosmic streams is expected to be particularly prominent and not yet significantly disrupted by halo-induced shocks); 4) observation of a high enough transition, CO J=4-3, which does not suffer from the reduced contrast against the Cosmic Microwave Background (CMB) (warmer by a factor of $\rm 1+z$), which may prevent the detection of very extended emission in lower CO transitions observations (\citealp{Zhang2016}).
\\\\
Our interpretation is further supported by the detection of gas rich systems on even larger scales, $\sim 550$ kpc, predominantly distributed in the same
direction as the central 40~kpc elongated structure (Sec.~\ref{sec:COlargescale}). The inferred mass of such  CO emitters is in the range of that estimated for gas rich proto-galaxies expected to form in cosmic filaments according to simulations (\citealp{Dekel2009}; \citealp{Nelson2016}). 
Therefore, we suggest that some of the detected gaseous systems (especially those with velocities within a few hundreds $\rm km \,s^{-1}$, consistent with them being gravitationally associated to the 40 kpc scale gaseous structure) may be tracing the same filamentary streams terminating onto the central massive galaxy and producing the inner elongated condensation of molecular gas. 
\\\\
Large molecular gas masses within filamentary accreting gas structures, traced by CO, indicate the presence of past and ongoing star formation in such streams, which has resulted in some pre-enrichment even before they enter the massive galaxies discs.
As mentioned in the introduction, such pre-enrichment of cold gaseous streams, due to gravitational collapse and star formation along the accreting filaments, eventually leading proto-galaxies formation, is expected by cosmic simulations (\citealp{Dekel2009}; \citealp{Nelson2016}; \citealp{Ceverino2016}; \citealp{Nelson2015}; \citealp{Darvish2014}). This scenario is further confirmed by the observation of ongoing star formation in the 40~kpc extended CO molecular structure and in the 550~kpc-scale CO systems which are associated with the stream (either through individual detections or stacking).
In this light, also the two optical companion galaxies of Candles-5001, may be the result of such fragmentation and star formation occurred along the filaments.\newline

One potential concern is that studies attempting to trace cold streams through absorption along the line of sight of quasars have pointed at very low metallicities \citep[e.g.][]{Fumagalli2016}. 
However, these studies have generally identified absorption systems around less massive galaxies. 
Moroever, even in massive galaxies, it is expected that the filling factor of dense (enriched) molecular systems along the streams is low, hence the chance of detecting them through random lines of sights probed by background quasars is very low.
It is however interesting to note that some studies, exploiting the analysis of absorption systems in the vicinity of high redshift galaxies, have indeed found evidence of gas in accretion with relatively high metallicities ($\rm \sim 0.4~Z_{\odot}$, \citealp{Bouche2013,Bouche2016}).

\section{Conclusions}

We have presented ALMA observations targeting the field of Candels-5001, the most massive galaxy  ($\rm M_{\star} =\rm 1.9 \times 10^{10}\,M_{\odot}$) in the redshift range $\rm 3<z<4$ within the GOODS-S field. 
This galaxy lies in an overdense region (located in a prominent redshift spike), likely tracing a forming proto-cluster (\citealp{Franck2016}). 
Taking advantage of such overdense environment, expected to more efficiently enrich the intergalactic environment, we investigated the molecular gas distribution traced by ALMA observations of the CO J=4-3 rotational transition. \newline
Our findings can be summarized as follows:
\begin{enumerate}
\item We detect a large structure of molecular gas reservoir around Candels-5001, extended over 40 kpc with a clearly elongated morphology (Fig.~\ref{fig:COflux_maps}).
\item By using a metallicity dependent CO-to-H$_2$ conversion factor (\citealp{Bolatto2013}) and a range of values for the CO J=4-3/CO J=1-0 lines flux ratio (\citealp{Carilli2013}), we infer for the gaseous structure a mass of $\rm M_{gas}\sim2-6 \times 10^{11} M_{\odot}$. 
About 60\% of the mass is not directly associated with either the central galaxy or its two satellites, but is distributed in the intergalactic medium in the vicinity of the central galaxy.
Such large amount of molecular mass in the CGM cannot be explained in terms of tidally stripped molecular gas due to gravitational interaction (see Sec.~\ref{sec:discussion}).
\item The molecular gas in the central 40~kpc has a complex kinematics.
The North-East external region of the gaseous structure appears to be characterized by two distinct kinematic components (Fig.~\ref{fig:kinematics}). The p-v diagram does not show any signature of coherent rotation of the large structure, instead  showing sub-structures with different velocities at the same location (Fig.~\ref{fig:pv_diagram}). As traced by the [OIII]5007$\rm \mathring{A}$ optical nebular lines, Candels-5001 and its two companions have individually resolved rotation curves. Altogether these kinematics evidences rule out the scenario in which the detected molecular structure is tracing a large rotating disc where the individual optical systems are sub-clumps.
\item We detect continuum thermal dust emission ($\rm \lambda_{rest} \sim 650\, \mu m$) associated with the CO structure on 40~kpc scales. Together with \textit{Herschel} data, the IR-to-mm continuum SED indicates a dust mass of about $1.6\times 10^9 M_{\odot}$. Using a dust-to-gas conversion factor appropriate for the metallicity in this systems, this dust mass corresponds to a gas mass of about $2\times 10^{11}~M_{\odot}$, consistent with that inferred from CO observations.
\item The extended continuum emission also suggests that the extended structure is undergoing star formation at a rate of about $\rm 0.1~M_{\odot}~yr^{-1}~kpc^{-2}$, or lower. This indicates that star formation in the extended structure is close to the Schmidt-Kennicutt relation (Fig.~\ref{fig:KS1}), but slightly less efficient than in normal star forming galaxies at this redshift.
\item We detect gas rich systems on scales of $\sim$ 500 kpc, within the ALMA field of view, most of which are distributed in the same NE-SW direction as the central CO $\sim$ 40 kpc elongated structure (Fig.~\ref{fig:largescale1}). Assuming metallicities similar to Candels-5001, and using the same calibrators discussed before, we infer for these systems gas masses of about $\rm 10^{10} - 10^{11}\, M_{\odot}$.
\item Some of the systems have individual ALMA continuum detections. The other systems are detected in the stacked ALMA continuum map and also in the stacked X-ray image. They are also marginally detected ($2.5\sigma$)  at optical wavelength in the stacked HST image (UV rest-frame).
Their inferred average SFR (about $\rm 30-120~M_{\odot}~yr^{-1}$), compared with their average content of molecular gas,
indicates that these systems are less efficient in forming stars than normal galaxies at this redshift (average depletion time-scales of 2~Gyr or
longer).
\item The velocity spread of the large scale CO systems, relative to Candels-5001,  is comparable with the velocity dispersions observed in other high-z protoclusters (\citealp{Casey2015}; \citealp{Chiang2015}), indicating that these systems may be tracing the gravitational field of the forming proto-cluster.
\end{enumerate}

We suggest that the 40~kpc extended CO emission, combined with the kinematics analysis, showing the presence of different kinematics components, is consistent with the system tracing the inner part of large scale accreting streams, feeding the central massive galaxy.
Our interpretation is corroborated by the CO emitting systems detected up to a distance of $\rm \sim 250\,kpc$ from the galaxy (in the ALMA field of view), whose distribution is aligned with the inner 40 kpc gaseous structure. 
Some of these objects, may be tracing the densest regions of the same filamentary streams terminating onto the massive galaxy.  \newline
The detected continuum thermal dust emission associated with the accreting structure on the 40~kpc scale, as well as the X-ray, optical and millimeter emission detected in the CO systems on large scales, indicate that a fraction of gas in the streams undergoes star formation (although inefficiently) even before accreting onto the central galaxy. 

These findings are in agreement with several cosmological models of galaxy formation and evolution at early epochs, according to which cosmic gaseous streams feeding massive galaxies are clumpy and may undergo gravitational collapse, by forming stars and protogalaxies, hence enriching a fraction of the gas in the streams, even before accreting onto the massive galaxy (\citealp{Dekel2009}; \citealp{SanchezAlmeida2014}; \citealp{Keres2009}; \citealp{Nelson2016}; \citealp{Ceverino2016}; \citealp{Pallottini2014}).

\section*{Acknowledgements}
This paper makes use of the following ALMA data: ADS/JAO.ALMA\#2012.1.00423.S. 
ALMA is a partnership of ESO (representing its member states), NSF (USA) and NINS (Japan), together with NRC (Canada) and NSC and ASIAA (Taiwan) and KASI (Republic of Korea), in cooperation with the Republic of Chile. 
The Joint ALMA Observatory is operated by ESO, AUI/NRAO and NAOJ.
RM acknowledge support from the ERC Advanced Grant 695671 \textquoteleft QUENCH\textquoteright.
RM, SC and FB acknowledge support from the Science and Technology Facilities Council (STFC). \newline
The research leading to these results has received funding from the 
European Research Council under the European Union's 
Seventh Framework Programme (FP/2007-2013) / ERC Grant Agreement n. 306476.


\bibliographystyle{mnras}

\bibliography{biblio}

\begin{thebibliography}{}
\makeatletter
\relax
\def\mn@urlcharsother{\let\do\@makeother \do\$\do\&\do\#\do\^\do\_\do\%\do\~}
\def\mn@doi{\begingroup\mn@urlcharsother \@ifnextchar [ {\mn@doi@}
  {\mn@doi@[]}}
\def\mn@doi@[#1]#2{\def\@tempa{#1}\ifx\@tempa\@empty \href
  {http://dx.doi.org/#2} {doi:#2}\else \href {http://dx.doi.org/#2} {#1}\fi
  \endgroup}
\def\mn@eprint#1#2{\mn@eprint@#1:#2::\@nil}
\def\mn@eprint@arXiv#1{\href {http://arxiv.org/abs/#1} {{\tt arXiv:#1}}}
\def\mn@eprint@dblp#1{\href {http://dblp.uni-trier.de/rec/bibtex/#1.xml}
  {dblp:#1}}
\def\mn@eprint@#1:#2:#3:#4\@nil{\def\@tempa {#1}\def\@tempb {#2}\def\@tempc
  {#3}\ifx \@tempc \@empty \let \@tempc \@tempb \let \@tempb \@tempa \fi \ifx
  \@tempb \@empty \def\@tempb {arXiv}\fi \@ifundefined
  {mn@eprint@\@tempb}{\@tempb:\@tempc}{\expandafter \expandafter \csname
  mn@eprint@\@tempb\endcsname \expandafter{\@tempc}}}

\bibitem[\protect\citeauthoryear{{Birnboim} \& {Dekel}}{{Birnboim} \&
  {Dekel}}{2003}]{Birnboim2003}
{Birnboim} Y.,  {Dekel} A.,  2003, \mn@doi [\mnras]
  {10.1046/j.1365-8711.2003.06955.x}, \href
  {http://adsabs.harvard.edu/abs/2003MNRAS.345..349B} {345, 349}

\bibitem[\protect\citeauthoryear{{Bolatto}, {Wolfire}  \& {Leroy}}{{Bolatto}
  et~al.}{2013}]{Bolatto2013}
{Bolatto} A.~D.,  {Wolfire} M.,   {Leroy} A.~K.,  2013, \mn@doi [\araa]
  {10.1146/annurev-astro-082812-140944}, \href
  {http://adsabs.harvard.edu/abs/2013ARA%26A..51..207B} {51, 207}

\bibitem[\protect\citeauthoryear{{Borisova} et~al.,}{{Borisova}
  et~al.}{2016}]{Borisova2016}
{Borisova} E.,  et~al., 2016, preprint, \href
  {http://adsabs.harvard.edu/abs/2016arXiv160501422B} {} (\mn@eprint {arXiv}
  {1605.01422})

\bibitem[\protect\citeauthoryear{{Bouch{\'e}}, {Murphy}, {Kacprzak},
  {P{\'e}roux}, {Contini}, {Martin}  \& {Dessauges-Zavadsky}}{{Bouch{\'e}}
  et~al.}{2013}]{Bouche2013}
{Bouch{\'e}} N.,  {Murphy} M.~T.,  {Kacprzak} G.~G.,  {P{\'e}roux} C.,
  {Contini} T.,  {Martin} C.~L.,   {Dessauges-Zavadsky} M.,  2013, \mn@doi
  [Science] {10.1126/science.1234209}, \href
  {http://adsabs.harvard.edu/abs/2013Sci...341...50B} {341, 50}

\bibitem[\protect\citeauthoryear{{Bouch{\'e}} et~al.,}{{Bouch{\'e}}
  et~al.}{2016}]{Bouche2016}
{Bouch{\'e}} N.,  et~al., 2016, \mn@doi [\apj] {10.3847/0004-637X/820/2/121},
  \href {http://adsabs.harvard.edu/abs/2016ApJ...820..121B} {820, 121}

\bibitem[\protect\citeauthoryear{{Brooks}, {Governato}, {Quinn}, {Brook}  \&
  {Wadsley}}{{Brooks} et~al.}{2009}]{Brooks2009}
{Brooks} A.~M.,  {Governato} F.,  {Quinn} T.,  {Brook} C.~B.,   {Wadsley} J.,
  2009, \mn@doi [\apj] {10.1088/0004-637X/694/1/396}, \href
  {http://adsabs.harvard.edu/abs/2009ApJ...694..396B} {694, 396}

\bibitem[\protect\citeauthoryear{{Bunker}, {Marleau}  \& {Graham}}{{Bunker}
  et~al.}{1998}]{Bunker1998}
{Bunker} A.~J.,  {Marleau} F.~R.,   {Graham} J.~R.,  1998, \mn@doi [\aj]
  {10.1086/300623}, \href {http://adsabs.harvard.edu/abs/1998AJ....116.2086B}
  {116, 2086}

\bibitem[\protect\citeauthoryear{{Burgarella}, {Buat}  \&
  {Iglesias-P{\'a}ramo}}{{Burgarella} et~al.}{2005}]{Burgarella2005}
{Burgarella} D.,  {Buat} V.,   {Iglesias-P{\'a}ramo} J.,  2005, \mn@doi
  [\mnras] {10.1111/j.1365-2966.2005.09131.x}, \href
  {http://adsabs.harvard.edu/abs/2005MNRAS.360.1413B} {360, 1413}

\bibitem[\protect\citeauthoryear{{Calzetti}, {Armus}, {Bohlin}, {Kinney},
  {Koornneef}  \& {Storchi-Bergmann}}{{Calzetti} et~al.}{2000}]{Calzetti2000}
{Calzetti} D.,  {Armus} L.,  {Bohlin} R.~C.,  {Kinney} A.~L.,  {Koornneef} J.,
   {Storchi-Bergmann} T.,  2000, \mn@doi [\apj] {10.1086/308692}, \href
  {http://adsabs.harvard.edu/abs/2000ApJ...533..682C} {533, 682}

\bibitem[\protect\citeauthoryear{{Cantalupo}, {Porciani}, {Lilly}  \&
  {Miniati}}{{Cantalupo} et~al.}{2005}]{Cantalupo2005}
{Cantalupo} S.,  {Porciani} C.,  {Lilly} S.~J.,   {Miniati} F.,  2005, \mn@doi
  [\apj] {10.1086/430758}, \href
  {http://adsabs.harvard.edu/abs/2005ApJ...628...61C} {628, 61}

\bibitem[\protect\citeauthoryear{{Cantalupo}, {Lilly}  \&
  {Haehnelt}}{{Cantalupo} et~al.}{2012}]{Cantalupo2012}
{Cantalupo} S.,  {Lilly} S.~J.,   {Haehnelt} M.~G.,  2012, \mn@doi [\mnras]
  {10.1111/j.1365-2966.2012.21529.x}, \href
  {http://adsabs.harvard.edu/abs/2012MNRAS.425.1992C} {425, 1992}

\bibitem[\protect\citeauthoryear{{Cantalupo}, {Arrigoni-Battaia}, {Prochaska},
  {Hennawi}  \& {Madau}}{{Cantalupo} et~al.}{2014}]{Cantalupo2014}
{Cantalupo} S.,  {Arrigoni-Battaia} F.,  {Prochaska} J.~X.,  {Hennawi} J.~F.,
  {Madau} P.,  2014, \mn@doi [\nat] {10.1038/nature12898}, \href
  {http://adsabs.harvard.edu/abs/2014Natur.506...63C} {506, 63}

\bibitem[\protect\citeauthoryear{{Carilli} \& {Walter}}{{Carilli} \&
  {Walter}}{2013}]{Carilli2013}
{Carilli} C.~L.,  {Walter} F.,  2013, \mn@doi [\araa]
  {10.1146/annurev-astro-082812-140953}, \href
  {http://adsabs.harvard.edu/abs/2013ARA%26A..51..105C} {51, 105}

\bibitem[\protect\citeauthoryear{{Casey} et~al.,}{{Casey}
  et~al.}{2015}]{Casey2015}
{Casey} C.~M.,  et~al., 2015, \mn@doi [\apjl] {10.1088/2041-8205/808/2/L33},
  \href {http://adsabs.harvard.edu/abs/2015ApJ...808L..33C} {808, L33}

\bibitem[\protect\citeauthoryear{{Cattaneo}, {Dekel}, {Devriendt}, {Guiderdoni}
   \& {Blaizot}}{{Cattaneo} et~al.}{2006}]{Cattaneo2006}
{Cattaneo} A.,  {Dekel} A.,  {Devriendt} J.,  {Guiderdoni} B.,   {Blaizot} J.,
  2006, \mn@doi [\mnras] {10.1111/j.1365-2966.2006.10608.x}, \href
  {http://adsabs.harvard.edu/abs/2006MNRAS.370.1651C} {370, 1651}

\bibitem[\protect\citeauthoryear{{Ceverino}, {S{\'a}nchez Almeida}, {Mu{\~n}oz
  Tu{\~n}{\'o}n}, {Dekel}, {Elmegreen}, {Elmegreen}  \& {Primack}}{{Ceverino}
  et~al.}{2016}]{Ceverino2016}
{Ceverino} D.,  {S{\'a}nchez Almeida} J.,  {Mu{\~n}oz Tu{\~n}{\'o}n} C.,
  {Dekel} A.,  {Elmegreen} B.~G.,  {Elmegreen} D.~M.,   {Primack} J.,  2016,
  \mn@doi [\mnras] {10.1093/mnras/stw064}, \href
  {http://adsabs.harvard.edu/abs/2016MNRAS.457.2605C} {457, 2605}

\bibitem[\protect\citeauthoryear{{Chiang} et~al.,}{{Chiang}
  et~al.}{2015}]{Chiang2015}
{Chiang} Y.-K.,  et~al., 2015, \mn@doi [\apj] {10.1088/0004-637X/808/1/37},
  \href {http://adsabs.harvard.edu/abs/2015ApJ...808...37C} {808, 37}

\bibitem[\protect\citeauthoryear{{Cresci}, {Mannucci}, {Maiolino}, {Marconi},
  {Gnerucci}  \& {Magrini}}{{Cresci} et~al.}{2010}]{Cresci2010}
{Cresci} G.,  {Mannucci} F.,  {Maiolino} R.,  {Marconi} A.,  {Gnerucci} A.,
  {Magrini} L.,  2010, \mn@doi [\nat] {10.1038/nature09451}, \href
  {http://adsabs.harvard.edu/abs/2010Natur.467..811C} {467, 811}

\bibitem[\protect\citeauthoryear{{Curti}, {Cresci}, {Mannucci}, {Marconi},
  {Maiolino}  \& {Esposito}}{{Curti} et~al.}{2016}]{Curti2016}
{Curti} M.,  {Cresci} G.,  {Mannucci} F.,  {Marconi} A.,  {Maiolino} R.,
  {Esposito} S.,  2016, preprint, \href
  {http://adsabs.harvard.edu/abs/2016arXiv161006939C} {} (\mn@eprint {arXiv}
  {1610.06939})

\bibitem[\protect\citeauthoryear{{Darvish}, {Sobral}, {Mobasher}, {Scoville},
  {Best}, {Sales}  \& {Smail}}{{Darvish} et~al.}{2014}]{Darvish2014}
{Darvish} B.,  {Sobral} D.,  {Mobasher} B.,  {Scoville} N.~Z.,  {Best} P.,
  {Sales} L.~V.,   {Smail} I.,  2014, \mn@doi [\apj]
  {10.1088/0004-637X/796/1/51}, \href
  {http://adsabs.harvard.edu/abs/2014ApJ...796...51D} {796, 51}

\bibitem[\protect\citeauthoryear{{Decarli} et~al.,}{{Decarli}
  et~al.}{2016}]{Decarli2016}
{Decarli} R.,  et~al., 2016, preprint, \href
  {http://adsabs.harvard.edu/abs/2016arXiv160706771D} {} (\mn@eprint {arXiv}
  {1607.06771})

\bibitem[\protect\citeauthoryear{{Dekel} \& {Birnboim}}{{Dekel} \&
  {Birnboim}}{2006}]{Dekel2006}
{Dekel} A.,  {Birnboim} Y.,  2006, \mn@doi [\mnras]
  {10.1111/j.1365-2966.2006.10145.x}, \href
  {http://adsabs.harvard.edu/abs/2006MNRAS.368....2D} {368, 2}

\bibitem[\protect\citeauthoryear{{Dekel} et~al.,}{{Dekel}
  et~al.}{2009}]{Dekel2009}
{Dekel} A.,  et~al., 2009, \mn@doi [\nat] {10.1038/nature07648}, \href
  {http://adsabs.harvard.edu/abs/2009Natur.457..451D} {457, 451}

\bibitem[\protect\citeauthoryear{{Dijkstra} \& {Loeb}}{{Dijkstra} \&
  {Loeb}}{2009}]{Dijkstra2009}
{Dijkstra} M.,  {Loeb} A.,  2009, \mn@doi [\mnras]
  {10.1111/j.1365-2966.2009.15533.x}, \href
  {http://adsabs.harvard.edu/abs/2009MNRAS.400.1109D} {400, 1109}

\bibitem[\protect\citeauthoryear{{Dunlop} et~al.,}{{Dunlop}
  et~al.}{2016}]{Dunlop2016}
{Dunlop} J.~S.,  et~al., 2016, preprint, \href
  {http://adsabs.harvard.edu/abs/2016arXiv160600227D} {} (\mn@eprint {arXiv}
  {1606.00227})

\bibitem[\protect\citeauthoryear{{Emonts} et~al.,}{{Emonts}
  et~al.}{2013}]{Emonts2013}
{Emonts} B.~H.~C.,  et~al., 2013, \mn@doi [\mnras] {10.1093/mnras/stt147},
  \href {http://adsabs.harvard.edu/abs/2013MNRAS.430.3465E} {430, 3465}

\bibitem[\protect\citeauthoryear{{Fall} \& {Efstathiou}}{{Fall} \&
  {Efstathiou}}{1980}]{Fall1980}
{Fall} S.~M.,  {Efstathiou} G.,  1980, \mn@doi [\mnras]
  {10.1093/mnras/193.2.189}, \href
  {http://adsabs.harvard.edu/abs/1980MNRAS.193..189F} {193, 189}

\bibitem[\protect\citeauthoryear{{Feldmann}, {Gnedin}  \&
  {Kravtsov}}{{Feldmann} et~al.}{2011}]{Feldmann2011}
{Feldmann} R.,  {Gnedin} N.~Y.,   {Kravtsov} A.~V.,  2011, \mn@doi [\apj]
  {10.1088/0004-637X/732/2/115}, \href
  {http://adsabs.harvard.edu/abs/2011ApJ...732..115F} {732, 115}

\bibitem[\protect\citeauthoryear{{Fiore} et~al.,}{{Fiore}
  et~al.}{2012}]{Fiore2012}
{Fiore} F.,  et~al., 2012, \mn@doi [\aap] {10.1051/0004-6361/201117581}, \href
  {http://adsabs.harvard.edu/abs/2012A%26A...537A..16F} {537, A16}

\bibitem[\protect\citeauthoryear{{Franck} \& {McGaugh}}{{Franck} \&
  {McGaugh}}{2016}]{Franck2016}
{Franck} J.~R.,  {McGaugh} S.~S.,  2016, \mn@doi [\apj]
  {10.3847/0004-637X/817/2/158}, \href
  {http://adsabs.harvard.edu/abs/2016ApJ...817..158F} {817, 158}

\bibitem[\protect\citeauthoryear{{Fraternali}}{{Fraternali}}{2014}]{Fraternali2014}
{Fraternali} F.,  2014, in {Feltzing} S.,  {Zhao} G.,  {Walton} N.~A.,
  {Whitelock} P.,  eds,  IAU Symposium Vol. 298, Setting the scene for Gaia and
  LAMOST. pp 228--239 (\mn@eprint {arXiv} {1310.2956}),
  \mn@doi{10.1017/S1743921313006418}

\bibitem[\protect\citeauthoryear{{Fumagalli}, {O'Meara}  \&
  {Prochaska}}{{Fumagalli} et~al.}{2016}]{Fumagalli2016}
{Fumagalli} M.,  {O'Meara} J.~M.,   {Prochaska} J.~X.,  2016, \mn@doi [\mnras]
  {10.1093/mnras/stv2616}, \href
  {http://adsabs.harvard.edu/abs/2016MNRAS.455.4100F} {455, 4100}

\bibitem[\protect\citeauthoryear{{Garnett}, {Shields}, {Skillman}, {Sagan}  \&
  {Dufour}}{{Garnett} et~al.}{1997}]{Garnett1997}
{Garnett} D.~R.,  {Shields} G.~A.,  {Skillman} E.~D.,  {Sagan} S.~P.,
  {Dufour} R.~J.,  1997, \apj, \href
  {http://adsabs.harvard.edu/abs/1997ApJ...489...63G} {489, 63}

\bibitem[\protect\citeauthoryear{{Geach} et~al.,}{{Geach}
  et~al.}{2009}]{Geach2009}
{Geach} J.~E.,  et~al., 2009, \mn@doi [\apj] {10.1088/0004-637X/700/1/1}, \href
  {http://adsabs.harvard.edu/abs/2009ApJ...700....1G} {700, 1}

\bibitem[\protect\citeauthoryear{{Genel} et~al.,}{{Genel}
  et~al.}{2012}]{Genel2012}
{Genel} S.,  et~al., 2012, \mn@doi [\apj] {10.1088/0004-637X/745/1/11}, \href
  {http://adsabs.harvard.edu/abs/2012ApJ...745...11G} {745, 11}

\bibitem[\protect\citeauthoryear{{Genzel} et~al.,}{{Genzel}
  et~al.}{2006}]{Genzel2006}
{Genzel} R.,  et~al., 2006, \mn@doi [\nat] {10.1038/nature05052}, \href
  {http://adsabs.harvard.edu/abs/2006Natur.442..786G} {442, 786}

\bibitem[\protect\citeauthoryear{{Genzel} et~al.,}{{Genzel}
  et~al.}{2010}]{Genzel2010}
{Genzel} R.,  et~al., 2010, \mn@doi [\mnras]
  {10.1111/j.1365-2966.2010.16969.x}, \href
  {http://adsabs.harvard.edu/abs/2010MNRAS.407.2091G} {407, 2091}

\bibitem[\protect\citeauthoryear{{Genzel} et~al.,}{{Genzel}
  et~al.}{2011}]{Genzel2011}
{Genzel} R.,  et~al., 2011, \mn@doi [\apj] {10.1088/0004-637X/733/2/101}, \href
  {http://adsabs.harvard.edu/abs/2011ApJ...733..101G} {733, 101}

\bibitem[\protect\citeauthoryear{{Genzel} et~al.,}{{Genzel}
  et~al.}{2014}]{Genzel2014}
{Genzel} R.,  et~al., 2014, \mn@doi [\apj] {10.1088/0004-637X/785/1/75}, \href
  {http://adsabs.harvard.edu/abs/2014ApJ...785...75G} {785, 75}

\bibitem[\protect\citeauthoryear{{Genzel} et~al.,}{{Genzel}
  et~al.}{2015}]{Genzel2015}
{Genzel} R.,  et~al., 2015, \mn@doi [\apj] {10.1088/0004-637X/800/1/20}, \href
  {http://adsabs.harvard.edu/abs/2015ApJ...800...20G} {800, 20}

\bibitem[\protect\citeauthoryear{{Goerdt}, {Dekel}, {Sternberg}, {Gnat}  \&
  {Ceverino}}{{Goerdt} et~al.}{2012}]{Goerdt2012}
{Goerdt} T.,  {Dekel} A.,  {Sternberg} A.,  {Gnat} O.,   {Ceverino} D.,  2012,
  \mn@doi [\mnras] {10.1111/j.1365-2966.2012.21397.x}, \href
  {http://adsabs.harvard.edu/abs/2012MNRAS.424.2292G} {424, 2292}

\bibitem[\protect\citeauthoryear{{Grazian} et~al.,}{{Grazian}
  et~al.}{2006}]{Grazian2006}
{Grazian} A.,  et~al., 2006, \mn@doi [\aap] {10.1051/0004-6361:20053979}, \href
  {http://adsabs.harvard.edu/abs/2006A%26A...449..951G} {449, 951}

\bibitem[\protect\citeauthoryear{{Gullberg} et~al.,}{{Gullberg}
  et~al.}{2016}]{Gullberg2016}
{Gullberg} B.,  et~al., 2016, \mn@doi [\aap] {10.1051/0004-6361/201527647},
  \href {http://adsabs.harvard.edu/abs/2016A%26A...591A..73G} {591, A73}

\bibitem[\protect\citeauthoryear{{Haardt} \& {Madau}}{{Haardt} \&
  {Madau}}{1996}]{Haardt1996}
{Haardt} F.,  {Madau} P.,  1996, \mn@doi [\apj] {10.1086/177035}, \href
  {http://adsabs.harvard.edu/abs/1996ApJ...461...20H} {461, 20}

\bibitem[\protect\citeauthoryear{{Hennawi}, {Prochaska}, {Cantalupo}  \&
  {Arrigoni-Battaia}}{{Hennawi} et~al.}{2015}]{Hennawi2015}
{Hennawi} J.~F.,  {Prochaska} J.~X.,  {Cantalupo} S.,   {Arrigoni-Battaia} F.,
  2015, \mn@doi [Science] {10.1126/science.aaa5397}, \href
  {http://adsabs.harvard.edu/abs/2015Sci...348..779H} {348, 779}

\bibitem[\protect\citeauthoryear{{Hodge}, {Carilli}, {Walter}, {de Blok},
  {Riechers}, {Daddi}  \& {Lentati}}{{Hodge} et~al.}{2012}]{Hodge2012}
{Hodge} J.~A.,  {Carilli} C.~L.,  {Walter} F.,  {de Blok} W.~J.~G.,  {Riechers}
  D.,  {Daddi} E.,   {Lentati} L.,  2012, \mn@doi [\apj]
  {10.1088/0004-637X/760/1/11}, \href
  {http://adsabs.harvard.edu/abs/2012ApJ...760...11H} {760, 11}

\bibitem[\protect\citeauthoryear{{Kere{\v s}}, {Katz}, {Weinberg}  \&
  {Dav{\'e}}}{{Kere{\v s}} et~al.}{2005}]{Keres2005}
{Kere{\v s}} D.,  {Katz} N.,  {Weinberg} D.~H.,   {Dav{\'e}} R.,  2005, \mn@doi
  [\mnras] {10.1111/j.1365-2966.2005.09451.x}, \href
  {http://adsabs.harvard.edu/abs/2005MNRAS.363....2K} {363, 2}

\bibitem[\protect\citeauthoryear{{Kere{\v s}}, {Katz}, {Fardal}, {Dav{\'e}}  \&
  {Weinberg}}{{Kere{\v s}} et~al.}{2009}]{Keres2009}
{Kere{\v s}} D.,  {Katz} N.,  {Fardal} M.,  {Dav{\'e}} R.,   {Weinberg} D.~H.,
  2009, \mn@doi [\mnras] {10.1111/j.1365-2966.2009.14541.x}, \href
  {http://adsabs.harvard.edu/abs/2009MNRAS.395..160K} {395, 160}

\bibitem[\protect\citeauthoryear{{Kollmeier}, {Zheng}, {Dav{\'e}}, {Gould},
  {Katz}, {Miralda-Escud{\'e}}  \& {Weinberg}}{{Kollmeier}
  et~al.}{2010}]{Kollmeier2010}
{Kollmeier} J.~A.,  {Zheng} Z.,  {Dav{\'e}} R.,  {Gould} A.,  {Katz} N.,
  {Miralda-Escud{\'e}} J.,   {Weinberg} D.~H.,  2010, \mn@doi [\apj]
  {10.1088/0004-637X/708/2/1048}, \href
  {http://adsabs.harvard.edu/abs/2010ApJ...708.1048K} {708, 1048}

\bibitem[\protect\citeauthoryear{{Leroy} et~al.,}{{Leroy}
  et~al.}{2011}]{Leroy2011}
{Leroy} A.~K.,  et~al., 2011, \mn@doi [\apj] {10.1088/0004-637X/737/1/12},
  \href {http://adsabs.harvard.edu/abs/2011ApJ...737...12L} {737, 12}

\bibitem[\protect\citeauthoryear{{Lowenthal}, {Hogan}, {Leach}, {Schmidt}  \&
  {Foltz}}{{Lowenthal} et~al.}{1990}]{Lowenthal1990}
{Lowenthal} J.~D.,  {Hogan} C.~J.,  {Leach} R.~W.,  {Schmidt} G.~D.,   {Foltz}
  C.~B.,  1990, \mn@doi [\apj] {10.1086/168884}, \href
  {http://adsabs.harvard.edu/abs/1990ApJ...357....3L} {357, 3}

\bibitem[\protect\citeauthoryear{{Magrini}, {Sestito}, {Randich}  \&
  {Galli}}{{Magrini} et~al.}{2009}]{Magrini2009}
{Magrini} L.,  {Sestito} P.,  {Randich} S.,   {Galli} D.,  2009, \mn@doi [\aap]
  {10.1051/0004-6361:200810634}, \href
  {http://adsabs.harvard.edu/abs/2009A%26A...494...95M} {494, 95}

\bibitem[\protect\citeauthoryear{{Maiolino} et~al.,}{{Maiolino}
  et~al.}{2008}]{Maiolino2008}
{Maiolino} R.,  et~al., 2008, \mn@doi [\aap] {10.1051/0004-6361:200809678},
  \href {http://adsabs.harvard.edu/abs/2008A%26A...488..463M} {488, 463}

\bibitem[\protect\citeauthoryear{{Maiolino} et~al.,}{{Maiolino}
  et~al.}{2015}]{Maiolino2015}
{Maiolino} R.,  et~al., 2015, \mn@doi [\mnras] {10.1093/mnras/stv1194}, \href
  {http://adsabs.harvard.edu/abs/2015MNRAS.452...54M} {452, 54}

\bibitem[\protect\citeauthoryear{{Matteucci}}{{Matteucci}}{2014}]{Matteucci2014}
{Matteucci} F.,  2014, \mn@doi [The Origin of the Galaxy and Local Group,
  Saas-Fee Advanced Course, Volume 37.~ISBN 978-3-642-41719-1.~Springer-Verlag
  Berlin Heidelberg, 2014, p.~145] {10.1007/978-3-642-41720-7_2}, \href
  {http://adsabs.harvard.edu/abs/2014SAAS...37..145M} {37, 145}

\bibitem[\protect\citeauthoryear{{Matteucci} \& {Francois}}{{Matteucci} \&
  {Francois}}{1989}]{Matteucci1989}
{Matteucci} F.,  {Francois} P.,  1989, \mn@doi [\mnras]
  {10.1093/mnras/239.3.885}, \href
  {http://adsabs.harvard.edu/abs/1989MNRAS.239..885M} {239, 885}

\bibitem[\protect\citeauthoryear{{McMullin}, {Waters}, {Schiebel}, {Young}  \&
  {Golap}}{{McMullin} et~al.}{2007}]{Macmullin2007}
{McMullin} J.~P.,  {Waters} B.,  {Schiebel} D.,  {Young} W.,   {Golap} K.,
  2007, in {Shaw} R.~A.,  {Hill} F.,   {Bell} D.~J.,  eds,  Astronomical
  Society of the Pacific Conference Series Vol. 376, Astronomical Data Analysis
  Software and Systems XVI. p.~127

\bibitem[\protect\citeauthoryear{{Miley} et~al.,}{{Miley}
  et~al.}{2006}]{Miley2006}
{Miley} G.~K.,  et~al., 2006, \mn@doi [\apjl] {10.1086/508534}, \href
  {http://adsabs.harvard.edu/abs/2006ApJ...650L..29M} {650, L29}

\bibitem[\protect\citeauthoryear{{Narayanan} et~al.,}{{Narayanan}
  et~al.}{2006}]{Narayanan2006}
{Narayanan} D.,  et~al., 2006, \mn@doi [\apjl] {10.1086/504846}, \href
  {http://adsabs.harvard.edu/abs/2006ApJ...642L.107N} {642, L107}

\bibitem[\protect\citeauthoryear{{Nelson}, {Genel}, {Vogelsberger}, {Springel},
  {Sijacki}, {Torrey}  \& {Hernquist}}{{Nelson} et~al.}{2015}]{Nelson2015}
{Nelson} D.,  {Genel} S.,  {Vogelsberger} M.,  {Springel} V.,  {Sijacki} D.,
  {Torrey} P.,   {Hernquist} L.,  2015, \mn@doi [\mnras]
  {10.1093/mnras/stv017}, \href
  {http://adsabs.harvard.edu/abs/2015MNRAS.448...59N} {448, 59}

\bibitem[\protect\citeauthoryear{{Nelson}, {Genel}, {Pillepich},
  {Vogelsberger}, {Springel}  \& {Hernquist}}{{Nelson}
  et~al.}{2016}]{Nelson2016}
{Nelson} D.,  {Genel} S.,  {Pillepich} A.,  {Vogelsberger} M.,  {Springel} V.,
   {Hernquist} L.,  2016, \mn@doi [\mnras] {10.1093/mnras/stw1191}, \href
  {http://adsabs.harvard.edu/abs/2016MNRAS.460.2881N} {460, 2881}

\bibitem[\protect\citeauthoryear{{Nesvadba}, {Lehnert}, {Eisenhauer},
  {Gilbert}, {Tecza}  \& {Abuter}}{{Nesvadba} et~al.}{2006}]{Nesvadba2006}
{Nesvadba} N.~P.~H.,  {Lehnert} M.~D.,  {Eisenhauer} F.,  {Gilbert} A.,
  {Tecza} M.,   {Abuter} R.,  2006, \mn@doi [\apj] {10.1086/507266}, \href
  {http://adsabs.harvard.edu/abs/2006ApJ...650..693N} {650, 693}

\bibitem[\protect\citeauthoryear{{Ocvirk}, {Pichon}  \& {Teyssier}}{{Ocvirk}
  et~al.}{2008}]{Ocvirk2008}
{Ocvirk} P.,  {Pichon} C.,   {Teyssier} R.,  2008, \mn@doi [\mnras]
  {10.1111/j.1365-2966.2008.13763.x}, \href
  {http://adsabs.harvard.edu/abs/2008MNRAS.390.1326O} {390, 1326}

\bibitem[\protect\citeauthoryear{{Pallottini}, {Gallerani}  \&
  {Ferrara}}{{Pallottini} et~al.}{2014}]{Pallottini2014}
{Pallottini} A.,  {Gallerani} S.,   {Ferrara} A.,  2014, \mn@doi [\mnras]
  {10.1093/mnrasl/slu126}, \href
  {http://adsabs.harvard.edu/abs/2014MNRAS.444L.105P} {444, L105}

\bibitem[\protect\citeauthoryear{{Pallottini}, {Ferrara}, {Gallerani},
  {Vallini}, {Maiolino}  \& {Salvadori}}{{Pallottini}
  et~al.}{2016}]{Pallottini2016}
{Pallottini} A.,  {Ferrara} A.,  {Gallerani} S.,  {Vallini} L.,  {Maiolino} R.,
    {Salvadori} S.,  2016, preprint, \href
  {http://adsabs.harvard.edu/abs/2016arXiv160901719P} {} (\mn@eprint {arXiv}
  {1609.01719})

\bibitem[\protect\citeauthoryear{{Patr{\'{\i}}cio} et~al.,}{{Patr{\'{\i}}cio}
  et~al.}{2016}]{Patricio2016}
{Patr{\'{\i}}cio} V.,  et~al., 2016, \mn@doi [\mnras] {10.1093/mnras/stv2859},
  \href {http://adsabs.harvard.edu/abs/2016MNRAS.456.4191P} {456, 4191}

\bibitem[\protect\citeauthoryear{{Pentericci}, {Kurk}, {Carilli}, {Harris},
  {Miley}  \& {R{\"o}ttgering}}{{Pentericci} et~al.}{2002}]{Pentericci2002}
{Pentericci} L.,  {Kurk} J.~D.,  {Carilli} C.~L.,  {Harris} D.~E.,  {Miley}
  G.~K.,   {R{\"o}ttgering} H.~J.~A.,  2002, \mn@doi [\aap]
  {10.1051/0004-6361:20021368}, \href
  {http://adsabs.harvard.edu/abs/2002A%26A...396..109P} {396, 109}

\bibitem[\protect\citeauthoryear{{Prochaska}, {Lau}  \& {Hennawi}}{{Prochaska}
  et~al.}{2014}]{Prochaska2014}
{Prochaska} J.~X.,  {Lau} M.~W.,   {Hennawi} J.~F.,  2014, \mn@doi [\apj]
  {10.1088/0004-637X/796/2/140}, \href
  {http://adsabs.harvard.edu/abs/2014ApJ...796..140P} {796, 140}

\bibitem[\protect\citeauthoryear{{Ranalli}, {Comastri}  \& {Setti}}{{Ranalli}
  et~al.}{2003}]{Ranalli2003}
{Ranalli} P.,  {Comastri} A.,   {Setti} G.,  2003, \mn@doi [\aap]
  {10.1051/0004-6361:20021600}, \href
  {http://adsabs.harvard.edu/abs/2003A%26A...399...39R} {399, 39}

\bibitem[\protect\citeauthoryear{{Rees} \& {Ostriker}}{{Rees} \&
  {Ostriker}}{1977}]{Rees1977}
{Rees} M.~J.,  {Ostriker} J.~P.,  1977, \mn@doi [\mnras]
  {10.1093/mnras/179.4.541}, \href
  {http://adsabs.harvard.edu/abs/1977MNRAS.179..541R} {179, 541}

\bibitem[\protect\citeauthoryear{{Saintonge} et~al.,}{{Saintonge}
  et~al.}{2013}]{Saintonge2013}
{Saintonge} A.,  et~al., 2013, \mn@doi [\apj] {10.1088/0004-637X/778/1/2},
  \href {http://adsabs.harvard.edu/abs/2013ApJ...778....2S} {778, 2}

\bibitem[\protect\citeauthoryear{{S{\'a}nchez Almeida}, {Elmegreen},
  {Mu{\~n}oz-Tu{\~n}{\'o}n}  \& {Elmegreen}}{{S{\'a}nchez Almeida}
  et~al.}{2014}]{SanchezAlmeida2014}
{S{\'a}nchez Almeida} J.,  {Elmegreen} B.~G.,  {Mu{\~n}oz-Tu{\~n}{\'o}n} C.,
  {Elmegreen} D.~M.,  2014, \mn@doi [\aapr] {10.1007/s00159-014-0071-1}, \href
  {http://adsabs.harvard.edu/abs/2014A%26ARv..22...71S} {22, 71}

\bibitem[\protect\citeauthoryear{{Sancisi}, {Fraternali}, {Oosterloo}  \& {van
  der Hulst}}{{Sancisi} et~al.}{2008}]{Sancisi2008}
{Sancisi} R.,  {Fraternali} F.,  {Oosterloo} T.,   {van der Hulst} T.,  2008,
  \mn@doi [\aapr] {10.1007/s00159-008-0010-0}, \href
  {http://adsabs.harvard.edu/abs/2008A%26ARv..15..189S} {15, 189}

\bibitem[\protect\citeauthoryear{{Sanders} \& {Mirabel}}{{Sanders} \&
  {Mirabel}}{1996}]{Sanders1996}
{Sanders} D.~B.,  {Mirabel} I.~F.,  1996, \mn@doi [\araa]
  {10.1146/annurev.astro.34.1.749}, \href
  {http://adsabs.harvard.edu/abs/1996ARA%26A..34..749S} {34, 749}

\bibitem[\protect\citeauthoryear{{Santini} et~al.,}{{Santini}
  et~al.}{2009}]{Santini2009}
{Santini} P.,  et~al., 2009, \mn@doi [\aap] {10.1051/0004-6361/200811434},
  \href {http://adsabs.harvard.edu/abs/2009A%26A...504..751S} {504, 751}

\bibitem[\protect\citeauthoryear{{Santini} et~al.,}{{Santini}
  et~al.}{2015}]{Santini2015}
{Santini} P.,  et~al., 2015, \mn@doi [\apj] {10.1088/0004-637X/801/2/97}, \href
  {http://adsabs.harvard.edu/abs/2015ApJ...801...97S} {801, 97}

\bibitem[\protect\citeauthoryear{{Sargent} et~al.,}{{Sargent}
  et~al.}{2014}]{Sargent2014}
{Sargent} M.~T.,  et~al., 2014, \mn@doi [\apj] {10.1088/0004-637X/793/1/19},
  \href {http://adsabs.harvard.edu/abs/2014ApJ...793...19S} {793, 19}

\bibitem[\protect\citeauthoryear{{Scoville} et~al.,}{{Scoville}
  et~al.}{2014}]{Scoville2014}
{Scoville} N.,  et~al., 2014, \mn@doi [\apj] {10.1088/0004-637X/783/2/84},
  \href {http://adsabs.harvard.edu/abs/2014ApJ...783...84S} {783, 84}

\bibitem[\protect\citeauthoryear{{Scoville} et~al.,}{{Scoville}
  et~al.}{2016}]{Scoville2016}
{Scoville} N.,  et~al., 2016, \mn@doi [\apj] {10.3847/0004-637X/820/2/83},
  \href {http://adsabs.harvard.edu/abs/2016ApJ...820...83S} {820, 83}

\bibitem[\protect\citeauthoryear{{Shen}, {Madau}, {Guedes}, {Mayer},
  {Prochaska}  \& {Wadsley}}{{Shen} et~al.}{2013}]{Shen2013}
{Shen} S.,  {Madau} P.,  {Guedes} J.,  {Mayer} L.,  {Prochaska} J.~X.,
  {Wadsley} J.,  2013, \mn@doi [\apj] {10.1088/0004-637X/765/2/89}, \href
  {http://adsabs.harvard.edu/abs/2013ApJ...765...89S} {765, 89}

\bibitem[\protect\citeauthoryear{{Silk} \& {Mamon}}{{Silk} \&
  {Mamon}}{2012}]{Silk2012}
{Silk} J.,  {Mamon} G.~A.,  2012, \mn@doi [Research in Astronomy and
  Astrophysics] {10.1088/1674-4527/12/8/004}, \href
  {http://adsabs.harvard.edu/abs/2012RAA....12..917S} {12, 917}

\bibitem[\protect\citeauthoryear{{Stewart}, {Kaufmann}, {Bullock}, {Barton},
  {Maller}, {Diemand}  \& {Wadsley}}{{Stewart} et~al.}{2011}]{Stewart2011}
{Stewart} K.~R.,  {Kaufmann} T.,  {Bullock} J.~S.,  {Barton} E.~J.,  {Maller}
  A.~H.,  {Diemand} J.,   {Wadsley} J.,  2011, \mn@doi [\apj]
  {10.1088/0004-637X/738/1/39}, \href
  {http://adsabs.harvard.edu/abs/2011ApJ...738...39S} {738, 39}

\bibitem[\protect\citeauthoryear{{Tacconi} et~al.,}{{Tacconi}
  et~al.}{2013}]{Tacconi2013}
{Tacconi} L.~J.,  et~al., 2013, \mn@doi [\apj] {10.1088/0004-637X/768/1/74},
  \href {http://adsabs.harvard.edu/abs/2013ApJ...768...74T} {768, 74}

\bibitem[\protect\citeauthoryear{{Tinsley}}{{Tinsley}}{1981}]{Tinsley1981}
{Tinsley} B.~M.,  1981, \mn@doi [\apj] {10.1086/159425}, \href
  {http://adsabs.harvard.edu/abs/1981ApJ...250..758T} {250, 758}

\bibitem[\protect\citeauthoryear{{Troncoso} et~al.,}{{Troncoso}
  et~al.}{2014}]{Troncoso2014}
{Troncoso} P.,  et~al., 2014, \mn@doi [\aap] {10.1051/0004-6361/201322099},
  \href {http://adsabs.harvard.edu/abs/2014A%26A...563A..58T} {563, A58}

\bibitem[\protect\citeauthoryear{{Ueda} et~al.,}{{Ueda}
  et~al.}{2014}]{Ueda2014}
{Ueda} J.,  et~al., 2014, \mn@doi [\apjs] {10.1088/0067-0049/214/1/1}, \href
  {http://adsabs.harvard.edu/abs/2014ApJS..214....1U} {214, 1}

\bibitem[\protect\citeauthoryear{{Vanzella} et~al.,}{{Vanzella}
  et~al.}{2016}]{Vanzella2016}
{Vanzella} E.,  et~al., 2016, preprint, \href
  {http://adsabs.harvard.edu/abs/2016arXiv160703112V} {} (\mn@eprint {arXiv}
  {1607.03112})

\bibitem[\protect\citeauthoryear{{Venemans} et~al.,}{{Venemans}
  et~al.}{2007}]{Venemans2007}
{Venemans} B.~P.,  et~al., 2007, \mn@doi [\aap] {10.1051/0004-6361:20053941},
  \href {http://adsabs.harvard.edu/abs/2007A%26A...461..823V} {461, 823}

\bibitem[\protect\citeauthoryear{{White} \& {Rees}}{{White} \&
  {Rees}}{1978}]{White1978}
{White} S.~D.~M.,  {Rees} M.~J.,  1978, \mn@doi [\mnras]
  {10.1093/mnras/183.3.341}, \href
  {http://adsabs.harvard.edu/abs/1978MNRAS.183..341W} {183, 341}

\bibitem[\protect\citeauthoryear{{Wisotzki} et~al.,}{{Wisotzki}
  et~al.}{2016}]{Wisotzki2016}
{Wisotzki} L.,  et~al., 2016, \mn@doi [\aap] {10.1051/0004-6361/201527384},
  \href {http://adsabs.harvard.edu/abs/2016A%26A...587A..98W} {587, A98}

\bibitem[\protect\citeauthoryear{{Wolfire}, {Hollenbach}  \& {McKee}}{{Wolfire}
  et~al.}{2010}]{Wolfire2010}
{Wolfire} M.~G.,  {Hollenbach} D.,   {McKee} C.~F.,  2010, \mn@doi [\apj]
  {10.1088/0004-637X/716/2/1191}, \href
  {http://adsabs.harvard.edu/abs/2010ApJ...716.1191W} {716, 1191}

\bibitem[\protect\citeauthoryear{{Yang}, {Zabludoff}, {Dav{\'e}}, {Eisenstein},
  {Pinto}, {Katz}, {Weinberg}  \& {Barton}}{{Yang} et~al.}{2006}]{Yang2006}
{Yang} Y.,  {Zabludoff} A.~I.,  {Dav{\'e}} R.,  {Eisenstein} D.~J.,  {Pinto}
  P.~A.,  {Katz} N.,  {Weinberg} D.~H.,   {Barton} E.~J.,  2006, \mn@doi [\apj]
  {10.1086/497898}, \href {http://adsabs.harvard.edu/abs/2006ApJ...640..539Y}
  {640, 539}

\bibitem[\protect\citeauthoryear{{Zhang}, {Papadopoulos}, {Ivison}, {Galametz},
  {Smith}  \& {Xilouris}}{{Zhang} et~al.}{2016}]{Zhang2016}
{Zhang} Z.-Y.,  {Papadopoulos} P.~P.,  {Ivison} R.~J.,  {Galametz} M.,  {Smith}
  M.~W.~L.,   {Xilouris} E.~M.,  2016, \mn@doi [Royal Society Open Science]
  {10.1098/rsos.160025}, \href
  {http://adsabs.harvard.edu/abs/2016RSOS....360025Z} {3, 160025}

\bibitem[\protect\citeauthoryear{{van de Voort}, {Schaye}, {Booth}, {Haas}  \&
  {Dalla Vecchia}}{{van de Voort} et~al.}{2011}]{vandeVoort2011}
{van de Voort} F.,  {Schaye} J.,  {Booth} C.~M.,  {Haas} M.~R.,   {Dalla
  Vecchia} C.,  2011, \mn@doi [\mnras] {10.1111/j.1365-2966.2011.18565.x},
  \href {http://adsabs.harvard.edu/abs/2011MNRAS.414.2458V} {414, 2458}

\makeatother
\end{thebibliography}

\appendix

\section{Astrometric alignment}\label{sec:appendix_offset}

\begin{figure}
	\centering
	\includegraphics[width=1\columnwidth]{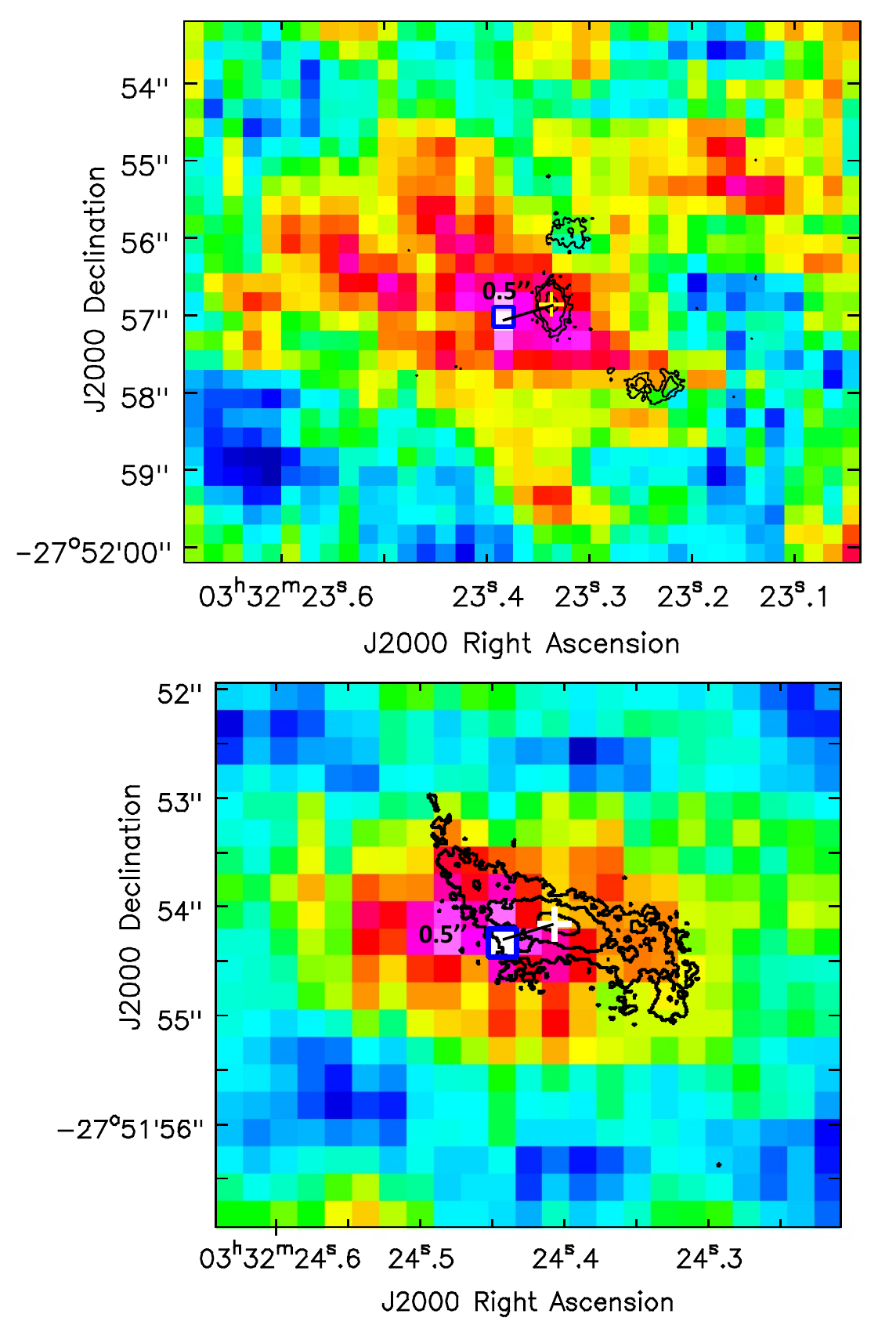}
	\caption{Astrometric shift. Top panel: optical HST image (contour) overlaid onto the ALMA continuum image (colour) of Candels-5001. Bottom panel: optical image of a foreground galaxy (COMBO17-25370, contours) overlaid onto its CO(2-1) emission serendipitously detected by ALMA (colour) within the field of view of Candels-5001. In both images the white cross indicates the peak of the optical emission while the blue square indicates the peak of the ALMA map. Clearly, in both cases there is a systemic  offset of 0.5$''$ between the optical and millimeter images in the same direction, indicative of an astrometric shift to be corrected.}
	\label{fig:shift}
\end{figure}

We have noticed that a systemic offset of about 0.5$''$ in the same direction is revealed between the optical image of Candels-5001 and the peak of the ALMA continuum emission, as well as between the image of a low redshift galaxy in the ALMA field of view (COMBO17-25370, MUSIC ID 04383) and the serendipitous detection of an emission line associated with the same galaxy (most likely CO J=2-1 at z=1.2). 
Fig.~\ref{fig:shift} shows the optical HST images of Candles-5001 (top) and of the serendipitous foreground galaxy COMBO17-25370 (bottom) overlay onto the ALMA continuum emission (top) and the CO J=2-1 transition of the serendipitous galaxy (bottom), illustrating that they are indeed subject to the same offset in the same direction.

As discussed in Sec.~\ref{sec:observations}, this kind of systemic offset have been identified in other studies comparing ALMA and optical data, and are ascribed to astrometric uncertainties associated with the optical images, or with the ALMA phase calibrator, or both of them.
As a consequence, we performed a slight astrometric alignment of $\sim$0.5$''$ in the NW direction, to match the ALMA data with the optical images. 
However, even if neglected, such alignment would not affect the conclusions of the paper.

\section{Multi-band images of Candels-5001}\label{sec:HST-Spitzer maps}
\begin{figure*}
	\centering
	\includegraphics[width=1\textwidth]{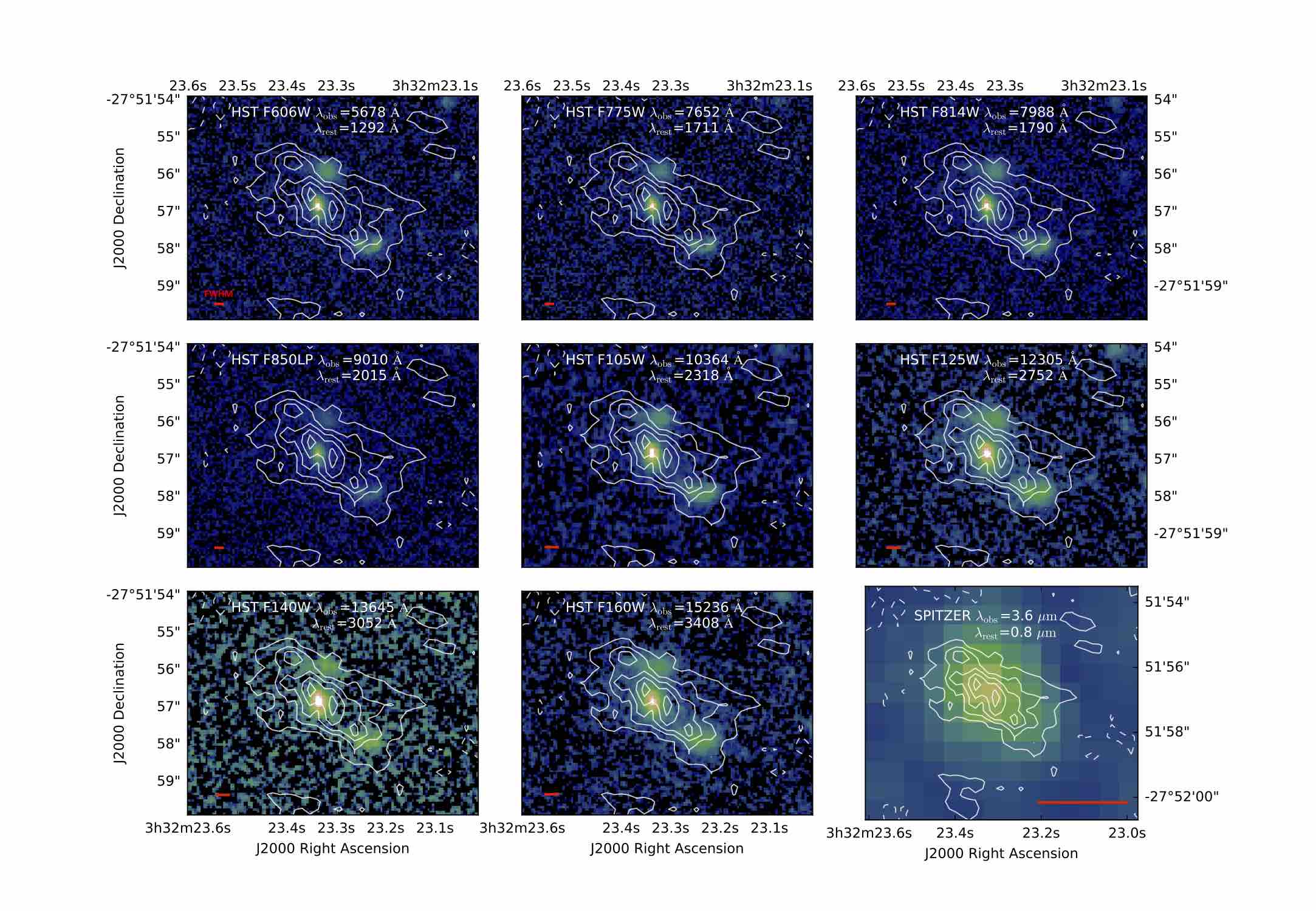}
	\caption{Overlay of CO emission with multiwavelenght images of Candels-5001. The CO(4-3) emission is shown with contours (as in ~\ref{fig:COflux_maps}),
		while the colour background images shows the available HST images (filter and rest-frame wavelength indicated within each panel) and the $\rm 3.6\,\mu m$ Spitzer image (bottom-right). The red bar in each panel indicates the size of the HST/Spitzer PSF in that band.}
	\label{fig:overlay}
\end{figure*}
In Fig.~\ref{fig:overlay} we show the overlay of the CO(4--3) map with all available HST images, as well as with the $\rm 3.6\,\mu m$ Spitzer image. The rest-frame wavelength
corresponding to each image is indicated in each panel. The tentative detection of some weak diffuse
emission in the HST F160W images, tracing continuum stellar emission at $\rm \lambda_{rest}\sim3400\rm \mathring{A}$, supports the scenario in which some dust-reddening is affecting the low surface brightness star formation occurring in this structure. 
The $\rm 3.6\,\mu m$ Spitzer image (Fig.~\ref{fig:overlay}) has too low (2.5$''$) resolution to investigate the extended emission, which could not be disentangled from the contribution associated with the two satellite galaxies.\newline

\section{Angular resolution effects Candels-5001}\label{sec:appendixsmoothing}
\begin{figure*}
	\centering
	\includegraphics[width=0.5\textwidth]{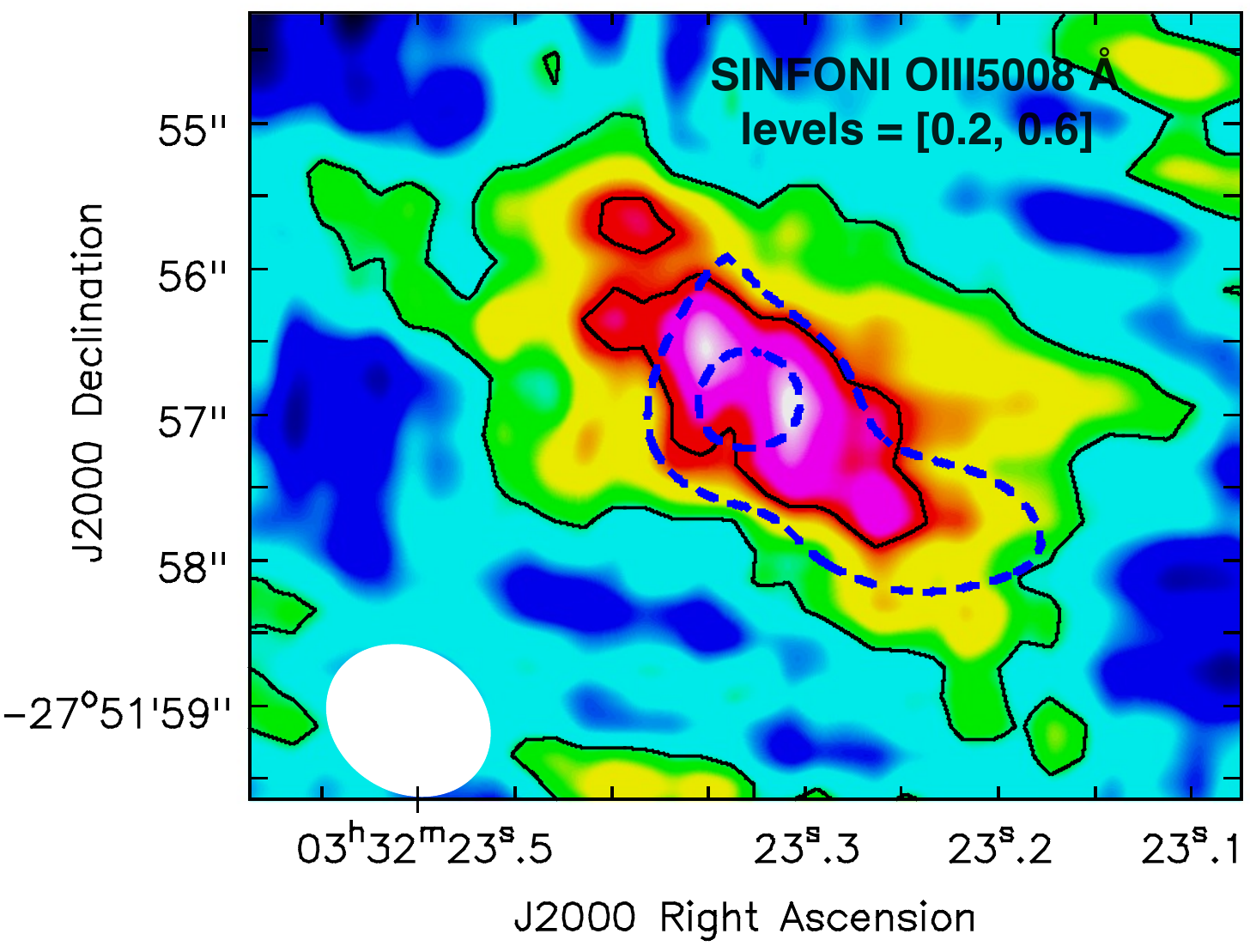}
	\caption{The background colour image is the CO map of the few tens kpc field around Candels-5001, obtained by ALMA. Black contours contain the 80\% and 40\% of the CO J=4-3 flux. The blue dashed contours contain the 80\% and 40\% of the [OIII] flux map obtained by smoothing the SINFONI/VLT data to the same resolution as the ALMA map.}
	\label{fig:smoothing}
\end{figure*}

In section Sec.~\ref{sec:extended} we claimed that the extended structure cannot be ascribed to molecular gas hosted in the three merging galaxies and \textquoteleft artificially\textquoteright\, smeared on a larger scale by the ALMA beam. 
We convolved the higher resolution [OIII]5007$\rm \mathring{A}$ SINFONI map (Fig.~\ref{fig:kinematics})  of Candels-5001 and its companions with the ALMA beam. 
In Fig.~\ref{fig:smoothing} we compare the spatial extension of the contours containing the 40\% and 80\% of the flux in the ALMA CO map with the same level contours of the SINFONI smoothed images, confirming that beam smearing of putative molecular gas in the two satellite galaxies (along with the molecular gas in Candles-5001) cannot account for the observed CO extent.

\section{Additional maps and tabulated properties of the CO emitting systems}\label{sec:largescale}

\begin{figure*}
	\centering
	\includegraphics[width=0.8\textwidth]{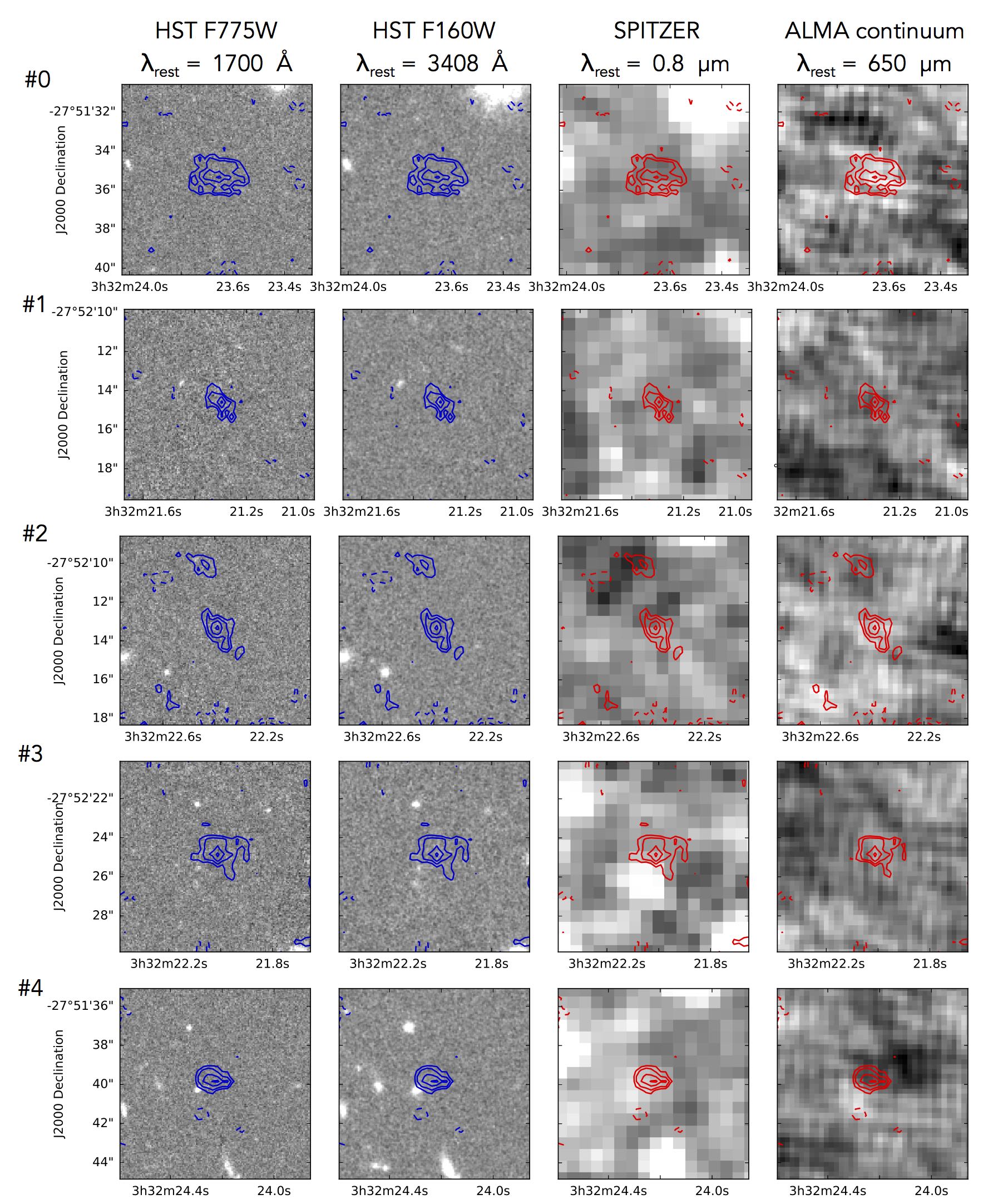}
	\caption{CO J=4-3 maps (contours) of the individual systems detected on large scales overlaid onto the HST, Spitzer and ALMA continuum images. Objects \#0 - \#4.}
	\label{fig:COsystems2_a}
\end{figure*}

Fig.~\ref{fig:COsystems2_a} shows the overlay of the CO emission observed in the systems detected on large scales with optical (F775W, HST), near-IR (F160W-HST and 3.6$\mu$m-Spitzer) and ALMA continuum images (see text for discussion). 
Table~\ref{tab:COsystems} lists some basic properties of the CO systems on large scales, where line fluxes have been corrected for the primary beam response function.

\begin{table*}
	\centering
	\caption{Summary of the properties of the CO systems detected on large scales.}
	\label{tab:COsystems}
	\begin{threeparttable}
		\begin{tabular}{ccccccc} % four columns, alignment for each
			\hline
			ID & Right Ascension & Declination & Line frequency & Line flux CO J=4-3 
			& Line width CO J=4-3 & Continuum flux\\
			&J2000 & J2000 & CO J=4-3 GHz & Jy $\rm km\,s^{-1}$ & Jy $\rm km\,s^{-1}$&
			$\rm \mu Jy$ \\
			\hline
			\#0& 03:32:23.70 &-27.51.33.8 & 102.72 & $0.110 \pm 0.022$& 125 & $24.1 \pm9.6$\tnote{a}\\ [0.2cm]
			\#1& 03:32:21.34 &-27.52.14.5 & 102.64 & $0.117 \pm 0.022$& 100 & $<9.6$\\ [0.2cm]
			\#2& 03:32:22.44 &-27.52.13.3 & 102.38 & $0.105 \pm 0.012$& 131 & $22.3 \pm9.6$\tnote{a}\\ [0.2cm]
			\#3& 03:32:22.06 &-27.52.24.8 & 103.10 & $0.202 \pm 0.038$& 214 & $<9.6$\\ [0.2cm]
			\#4& 03:32:24.27 &-27.51.39.8 & 103.05 & $0.104 \pm 0.018$& 110 & $<9.6$\\ [0.2cm]
			\#5& 03:32:24.70 &-27.52.11.5 & 102.85 & $0.107 \pm 0.013$& 125 & $33.4 \pm9.6$\\ [0.2cm]
			\#6& 03:32:24.61 &-27.52.10.3 & 103.85 & $0.102 \pm 0.018$& 83 & $<9.6$\\ [0.2cm]
			\#7& 03:32:21.82 &-27.51.58.5 & 103.75 & $0.136 \pm 0.024$& 289 & $<9.6$\\ [0.2cm]
			\#8& 03:32:24.55 &-27.51.30.3 & 103.83 & $0.116 \pm 0.021$& 69 & $<9.6$\\ [0.2cm]
			\hline
		\end{tabular}
		\begin{tablenotes}
			\item[a]  Marginal detections ($\sim2.5\sigma$)
		\end{tablenotes}
	\end{threeparttable}
	
\end{table*}

\begin{figure*}
	\centering
	\includegraphics[width=0.8\textwidth]{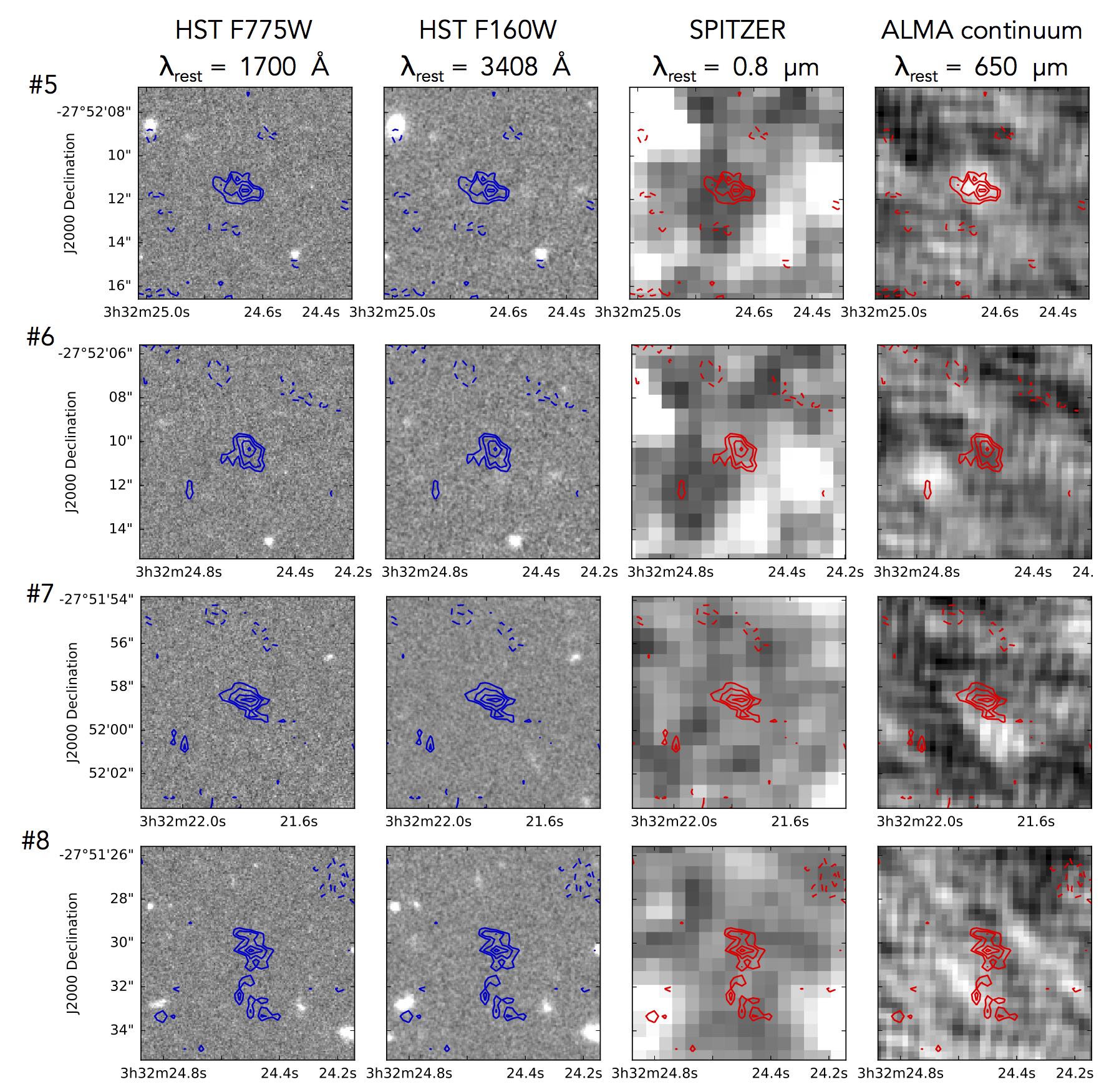}
	\caption{CO J=4-3 maps (contours) of the individual systems detected on large scales overlaid onto the HST, Spitzer and ALMA continuum images. Objects \#5 - \#8.}
	\label{fig:COsystems2_b}
\end{figure*}

% Don't change these lines
\bsp	% typesetting comment
\label{lastpage}

\end{document}